\documentclass[10pt,aps,prd,floats,floatfix,showpacs,superscriptaddress,nofootinbib]{revtex4-2}
\usepackage[utf8]{inputenc} 
\usepackage{graphicx,mathtools,amssymb,amsmath,amsthm,amsfonts,epsfig,epsf,bm,multirow}
\usepackage[outdir=./]{epstopdf}
\usepackage[usenames]{color}
\usepackage{tensor}
\usepackage{csquotes}
\usepackage{mathrsfs}
\usepackage{autobreak}
\usepackage{physics}
\usepackage{cancel}
\usepackage[font=small,labelfont=bf, justification=Justified,format=plain]{caption}
\usepackage{slashed}
\usepackage{pifont}
\usepackage[dvipsnames]{xcolor}
\usepackage[colorlinks=true]{hyperref}
\usepackage{cleveref}
\usepackage{siunitx}
\usepackage{soul}

\counterwithin*{equation}{section}

\def\del{\partial}

\def\be{\begin{equation}}
\def\ee{\end{equation}}
\def\bea{\begin{eqnarray}}
\def\eea{\end{eqnarray}}

\usepackage{xcolor}
\usepackage{amsmath}
\usepackage{cancel}
\usepackage{changes}

\definechangesauthor[name={Tony}, color=red]{AP}

\DeclareRobustCommand{\tonyin}[1]{%
  \ifmmode
    {\color{red}#1}%
  \else
    \added[id=AP]{#1}%
  \fi
}

\DeclareRobustCommand{\tonyout}[1]{%
  \ifmmode
    {\color{red}\cancel{#1}}%
  \else
    \deleted[id=AP]{#1}%
  \fi
}

\begin{document}

\title{A Lapse in the Cosmological Constant Problem with Bulk Dynamics}
\author{Justin Khoury}
\email{jkhoury@upenn.edu}
\affiliation{Center for Particle Cosmology, Department of Physics and Astronomy, University of Pennsylvania, 209 South 33rd St, Philadelphia, PA 19104, USA}

\author{Benjamin Muntz}
\email{benjamin.muntz@nottingham.ac.uk}
\affiliation{Nottingham Centre of Gravity, University of Nottingham,
University Park, Nottingham NG7 2RD, United Kingdom}
\affiliation{School of Physics and Astronomy, University of Nottingham, University Park, Nottingham NG7 2RD, United Kingdom}

\author{Antonio Padilla}
\email{antonio.padilla@nottingham.ac.uk}
\affiliation{Nottingham Centre of Gravity, University of Nottingham,
University Park, Nottingham NG7 2RD, United Kingdom}
\affiliation{School of Physics and Astronomy, University of Nottingham, University Park, Nottingham NG7 2RD, United Kingdom}


\begin{abstract}
We have previously proposed a new approach to the cosmological constant problem based on anisotropic scaling in a compact extra dimension. In the ultra-local~$z=0$ limit, the interplay between a projectable lapse, foliation-preserving diffeomorphisms, and higher-form fluxes renders the gravitational field equations insensitive to radiative corrections to the matter vacuum energy. Here we extend this framework beyond the ultra-local limit by introducing~$z=1$ deformations that restore dynamics along the compact direction. We show that vacuum-energy cancellation persists at the level of the background equations, while generic extrinsic-curvature couplings propagate an additional scalar ghost. A healthy quadratic spectrum selects the Fierz–Pauli relation between these couplings. This motivates a particularly simple realization in terms of five-dimensional Einstein gravity, a top-form flux, and a spacelike khoron scalar field that dynamically defines a preferred projectable foliation. In this covariant formulation, the global constraint arises from an auxiliary einbein on the space of khoron leaves, and constant shifts of the renormalized matter vacuum energy cancel algebraically from the intrinsic Einstein equations. Finally, allowing matter to propagate around the compact dimension generates finite, topology-sensitive Casimir contributions that are not automatically removed by the mechanism. Suppressing these contributions requires additional spectral conditions on the propagating bulk degrees of freedom.
\end{abstract}

\maketitle

\section{Introduction}

In~\cite{Khoury:2026eqr} we proposed a new approach to the cosmological constant problem based on anisotropic scaling in an extra dimension. In the deep infrared limit, corresponding to a~$z=0$ anisotropic scaling along the extra dimension, the theory becomes ultra-local in the compact direction, and the radiatively unstable matter vacuum energy decouples from the gravitational field equations. The mechanism relies on the interplay between foliation-preserving diffeomorphisms, a projectable lapse, and higher-form fluxes, leading to effective gravitational field equations that are insensitive to vacuum energy renormalization.

In this paper we extend that framework beyond the ultra-local limit by introducing~$z=1$ deformations that couple neighbouring slices of spacetime along the extra dimension. This allows us to test whether the vacuum-energy cancellation mechanism survives once dynamics along the compact direction are restored, and to investigate the resulting perturbative spectrum. Our main conclusion is that~$z=1$ deformations preserve vacuum-energy cancellation at the level of the background equations, but generic extrinsic-curvature couplings propagate an additional scalar ghost; a healthy quadratic spectrum selects the Fierz–Pauli relation between these couplings. This motivates a particularly simple realization in which gravity and the top-form sector are fully five-dimensional and diffeomorphism covariant, while a spacelike khoron dynamically enforces the preferred projectable foliation. Finally, allowing matter itself to propagate around the compact dimension generates finite Casimir energies that are not automatically cancelled, thereby separating the local vacuum-energy problem from topology-sensitive finite quantum effects.

In Sec.~\ref{sec:review}, we briefly review the~$z=0$ limit of the framework introduced in~\cite{Khoury:2026eqr}. The construction is based on a preferred foliation along the spatial extra dimension. This breaks five-dimensional diffeomorphism invariance down to foliation-preserving diffeomorphisms, allowing a projectable `lapse function' that depends only on the extra dimension and gives rise to a Hamiltonian constraint that is global, rather than local, along the four external spacetime directions. The mechanism is reminiscent of the global constraint in vacuum-energy sequestering~\cite{Kaloper:2013zca,Kaloper:2014dqa,Kaloper:2014fca,Kaloper:2015jra,Kaloper:2016yfa,Kaloper:2016jsd,DAmico:2017ngr,Padilla:2018hvp,Coltman:2019mql,El-Menoufi:2019qva}, which ensures that radiative corrections to the cosmological constant are cancelled. The details, however, are different: in the present construction the global character of the constraint follows directly from the underlying foliation symmetry and projectability condition.

Coupling the theory to a three-form field on each four-dimensional slice allows the resulting flux contribution to dynamically cancel the radiatively unstable matter vacuum energy. After integrating out the flux, the effective Einstein equations depend only on dynamical, non-vacuum matter contributions, while the matter vacuum energy drops out identically. The section also introduces the averaging procedures over spacetime slices and the compact direction that will play an important role once~$z=1$ couplings between slices are restored.

In Sec.~\ref{sec:def}, we extend the derivative expansion beyond the ultra-local limit by introducing~$z=1$ deformations that restore dynamics along the compact extra dimension through extrinsic-curvature and higher-form kinetic terms. The resulting theory couples neighbouring slices while preserving the global structure associated with the projectable lapse. The four-dimensional three-form is correspondingly lifted to a five-dimensional four-form, whose leading normal-derivative interactions preserve the foliation symmetry. After deriving the modified field equations and global constraints, we integrate out the flux sector and show that the matter vacuum energy again cancels from the effective Einstein equations. For maximally symmetric vacuum configurations, averaging over the compact direction together with periodicity enforces vanishing four-dimensional curvature.

Having shown that~$z=1$ deformations preserve vacuum-energy cancellation at the level of the background equations, we next turn to their perturbative stability. In Sec.~\ref{sec:lin} we study linearised fluctuations about the simplest vacuum configuration and examine the spectrum of gravitational modes arising from dynamics along the compact extra dimension. Upon Fourier expanding around the periodic direction, the theory gives rise to a tower of massive graviton modes whose mass terms are controlled by the extrinsic-curvature couplings introduced in the~$z=1$ theory. For generic parameter choices, these massive gravitons propagate an additional scalar ghost, reflecting the well-known pathology of non–Fierz–Pauli mass terms~\cite{Fierz:1939ix}. Requiring a healthy quadratic spectrum therefore selects the Fierz–Pauli relation between the two extrinsic-curvature terms. Five-dimensional Einstein gravity lies on this healthy branch and provides a particularly natural covariant realization. Importantly, vacuum-energy cancellation follows from projectability and is independent of this perturbative consistency condition.

The perturbative consistency condition motivates us to reconsider the mechanism in a particularly simple setting in which gravity and the top-form sector are fully five-dimensional and diffeomorphism covariant, while additional fields dynamically select a preferred foliation and enforce projectability. Sec.~\ref{sec:stuck} therefore presents a standalone realization formulated directly in terms of five-dimensional General Relativity coupled to a top-form flux and a spacelike ``khoron'' scalar that dynamically defines the preferred foliation. Projectability is enforced covariantly through an auxiliary einbein on the one-dimensional space of khoron leaves. Its equation of motion provides the global constraint that, together with the normal Einstein equation and the constant five-form flux, ensures that shifts of the renormalized matter vacuum energy drop out of the intrinsic gravitational equations. This formulation provides a covariant counterpart of the original mechanism. 

Finally, in Sec.~\ref{Casimir sec} we relax the assumption that matter is ultra-local along the compact direction. The zero-winding ultraviolet contributions remain local from the five-dimensional perspective and extensive in the proper circumference, and therefore retain the lapse dependence required for vacuum-energy cancellation. Non-zero winding contributions are qualitatively different: matter propagation around the compact dimension generates Casimir energies with non-trivial dependence on its circumference, which are not automatically removed by the global constraint. We derive these contributions explicitly and identify a spectral condition under which the leading Casimir contribution cancels.

\section{A recap of our solution to the cosmological constant problem with anisotropic scaling} \label{sec:review}

In this section, we review our previous work~\cite{Khoury:2026eqr}, in which radiative corrections to vacuum energy are dynamically cancelled thanks to anisotropic scaling in a compact extra dimension.
In this framework, Lorentz invariance holds exactly in the four extended spacetime dimensions, compatible with stringent experimental constraints~\cite{Sanner:2018atx}, but is explicitly broken along the extra dimension. 

Writing the five-dimensional coordinates as $(x^\mu,y)$, we assign anisotropic scaling weights according to
\begin{equation}\label{aniso_review}
x^\mu \to \ell^z x^\mu\,,
\qquad
y \to \ell y\,.
\end{equation}
This assignment provides a convenient way to organise the derivative expansion and classify operators according to their scaling dimension. It is inspired by the anisotropic scaling that appears in Lifshitz-type theories, originally introduced in the context of critical phenomena~\cite{Lifshitz:1941a,Lifshitz:1941b,Hornreich:1975} and later studied in gravitational settings by Hořava~\cite{Horava:2009uw,Horava:2009if}. For example, the free scalar theory
\[
S=\int \dd y\,\dd^4x\,\phi\left(\Box_4+\partial_y^{2z}\right)\phi
\]
is invariant under the scaling~\eqref{aniso_review}. In the present work, however, Eq.~\eqref{aniso_review} should be viewed primarily as a bookkeeping device for the anisotropic derivative expansion rather than as an exact symmetry of the complete theory. In particular, $z=1$ corresponds to isotropic scaling and permits the Lorentz-invariant kinetic operator $\Box_5$, but neither implies nor is implied by Lorentz invariance of the full interacting theory. Likewise, we do not require the matter sector or the full gravitational theory to display exact scale invariance. Meanwhile, for~$z = 0$, all~$y$-derivatives are absent, and the theory becomes ultra-local in~$y$.

To implement this structure gravitationally, the five-dimensional geometry is decomposed into four-dimensional hypersurfaces~$\Sigma_y$ at fixed~$y$, assuming only foliation-preserving diffeomorphisms,
\begin{equation}\label{fdiffs_review}
  y \rightarrow y - \xi(y)\,, 
  \qquad 
  x^\mu \rightarrow x^\mu - \xi^\mu(x,y)\,,
\end{equation}
as in Hořava--Lifshitz gravity~\cite{Horava:2009uw,Horava:2009if}.
The metric is written in ADM form as
\begin{equation} \label{metric}
  \dd s_5^2 
  = N^2 \dd y^2 
  + g_{\mu\nu}\big(\dd x^\mu + N^\mu \dd y\big)\big(\dd x^\nu + N^\nu \dd y\big)\,,
\end{equation}
where the lapse function~$N(y)$ depends only on the slicing coordinate (projectable limit),~$N^\mu(x,y)$ is the shift, and~$g_{\mu\nu}(x,y)$ is the induced four-dimensional metric. The projectability of the lapse is crucial: variation with respect to~$N$ produces a global constraint rather than a local equation. With some foresight we assume that the~$y$-direction is compact.

At leading order in $y$-derivatives (alternatively, after taking a~$z=0$ limit), a general action compatible with the symmetries~\eqref{fdiffs_review},
with matter loops included at fixed background metric, takes the form
\begin{equation}\label{eq:review_action}
  S = \frac{1}{2\kappa^2} 
  \int \dd y\, N(y) 
  \int_{\Sigma_y} \dd ^4x \sqrt{-g}\, R - \frac{1}{2\kappa^2}\int \dd y\, N(y) \int_{\Sigma_y} F_4 \wedge \star_4 F_4
  + \int \dd y\, N(y)\, \Gamma_{\rm m}[g_y,\Psi_y]
  + \text{boundary terms}\,,
\end{equation}
where~$\kappa^2 = 1/M_5^3$ in terms of the 5d Planck mass.\footnote{One could add a bare five-dimensional cosmological constant $\Lambda_5$ to the action. This, however, does not alter the mechanism presented here, since it can be effectively absorbed into the contribution from four-form fluxes.} Here~$\Gamma_{\rm m}$ is the four-dimensional matter contribution to the renormalized fixed-background effective action, obtained by integrating out matter fluctuations while keeping the ADM variables $(N,N^\mu,g_{\mu\nu})$ fixed. The notation $g_y$ and $\Psi_y$ denote, respectively, the metric and background matter fields restricted to the slice $\Sigma_y$.\footnote{We assume a regularization and renormalization prescription that does not violate the foliation-preserving diffeomorphisms. In the~$z=0$ limit the matter sector contains no
$y$-derivatives, so the matter path integral is local in $y$, although $\Gamma_{\rm m}$ may be non-local on each four-dimensional slice. Local
counterterms generated by matter loops therefore appear with the invariant measure $\int \dd y\,N(y)\int_{\Sigma_y}\dd^4x\sqrt{-g}\,{\cal O}_4[g_y,\Psi_y]$.}
Apart from the volume term isolated later in Eq.~\eqref{Gam split z=0}, purely geometric terms generated by matter loops, including local curvature counterterms and non-local
curvature form factors, are assigned to the renormalized gravitational action and are not included in~$\Gamma_{\rm m}$.
Notice moreover that matter fields are minimally coupled on each slice and smeared along the extra dimension. This places an upper bound on the proper length of the extra dimension from collider experiments,~$L \lesssim \SI{e-20}{\metre}$ \cite{ParticleDataGroup:2024cfk}. We have also introduced a three-form field confined to the four-dimensional hypersurfaces,
\begin{equation}\label{eq:A3definition}
  A_3(x,y) \equiv \frac{1}{3!}A_{\mu_1\mu_2\mu_3}(x,y)\,
  \theta^{\mu_1}\wedge\theta^{\mu_2}\wedge\theta^{\mu_3}\,,
  \qquad
  \theta^\mu = \dd x^\mu + N^\mu \dd y\,,
\end{equation}
with field strength~$F_4 = \dd _4 A_3$.
The corresponding equation of motion implies
\begin{equation}
  \star_4 F_4 = Q(y) \,,
\end{equation}
so that the four-form flux is constant on each slice but may vary along the extra dimension.

A key ingredient concerns the variational principle and boundary conditions. Along possible boundaries of the four-dimensional spacetime,~$\partial\Sigma_y$, the gravitational sector includes the Gibbons--Hawking--York term~\cite{Gibbons:1976ue}, while the three-form sector includes the Duncan--Jensen boundary term~\cite{Duncan:1989ug},
\begin{equation}\label{DJ_review}
  \frac{1}{\kappa^2}\int \dd y\, N \int_{\partial\Sigma_y} A_3\wedge \star_4 F_4\,,
\end{equation}
which enforces Neumann boundary conditions for the three-form. Since the lapse depends only on~$y$, imposing Dirichlet boundary conditions on~$N$ would eliminate bulk dynamics. Instead, we impose Neumann boundary conditions on the lapse. Consistency of the variational principle then requires~\cite{Kaloper:2016yfa}
\begin{equation}\label{bc_review}
  N\delta \gamma_{ij} = - \gamma_{ij}\delta N\,,
\end{equation}
which is equivalent to Dirichlet boundary conditions for the Einstein frame metric~$\tilde g_{\mu\nu}=N g_{\mu\nu}$. This ensures that the lapse function remains dynamical and leads to a non-trivial global constraint, as we will see.

Varying the action with respect to the metric yields
\begin{equation}\label{geqn_review}
  G_{\mu\nu}
  = -\frac{1}{2} Q^2 g_{\mu\nu}
  + \kappa^2 T^{(\text{m})}_{\mu\nu}\,,
\end{equation}
where~ $T^\text{(m)}_{\mu\nu}=-\frac{2}{\sqrt{-g}}\frac{\delta \Gamma_{\rm m}}{\delta g^{\mu\nu}}$ is the energy--momentum tensor for the matter sector. Variation with respect to the lapse gives the global constraint
\begin{equation}\label{Neqn_review}
    \int \dd ^4x \sqrt{-g}
  \left[
    \frac{1}{2} R
    -\frac{1}{2} Q^2(y)  \right]
    + \kappa^2 \Gamma_{\rm m}[g_y,\Psi_y]
  = 0\,.
\end{equation}
Because~$N$ is projectable, this is a spacetime-integrated condition rather than a local equation. The structure is therefore reminiscent of vacuum energy sequestering~\cite{Kaloper:2013zca,Kaloper:2014dqa,Kaloper:2014fca,Kaloper:2015jra,Kaloper:2016yfa,Kaloper:2016jsd,DAmico:2017ngr,Padilla:2018hvp,Coltman:2019mql,El-Menoufi:2019qva}, although here the global constraint emerges geometrically from the compact extra dimension rather than from explicit global variables.

To make the global quantities well defined, we first introduce a finite infrared region on each slice. Let $\Sigma_y^\circ\subseteq\Sigma_y$ be a
compact spacetime subregion and define its volume
\begin{equation}
\mathcal{V}_{4}(y)
=
\int_{\Sigma_y^\circ} \dd^4x\,\sqrt{-g}\, .
\label{volume def}
\end{equation}
If $\Sigma_y$ is compact, or more generally has finite four-volume, one simply takes $\Sigma_y^\circ=\Sigma_y$ and spacetime averages are well-defined without the need of an infrared regulator.
For $\Sigma_y$ non-compact, the non-regularized spacetime averages can be understood, when they exist, as the limit obtained by enlarging $\Sigma^\circ_y$ to cover the entirety (closure) of $\Sigma_y$. We hence think of the compact subset as labelled by some parameter $r$ where in the limit $r\to \infty$ we recover the entire spacetime slice. In what follows, we suppress this regulator label $r$ unless it is needed explicitly. Importantly, as we will see, the cancellation of the vacuum energy term occurs before taking the limit, and as such our mechanism is insensitive to the choice of regulator.

Substituting the trace of~\eqref{geqn_review} into the global constraint then gives
\begin{equation}
Q^2(y) = \kappa^2\left( \big\langle T^{(\text{m})} \big\rangle - 2\,\frac{\Gamma_{\rm m}}{\mathcal{V}_4} \right)\,,
\end{equation}
where
\begin{equation}
  \langle \mathcal{O} \rangle
  \equiv \frac{1}{\mathcal{V}_4} \int_{\Sigma_y^\circ} \dd^4x \sqrt{-g}\, \mathcal{O}(x,y)
\label{O avg}
\end{equation}
denotes the regulated spacetime average on each slice. Later we will also make use of averages over the compact dimension,
\begin{equation} \label{weightedav}
  \langle \mathcal{O} \rangle_{S^1}
  \equiv \frac{\int_{S^1} \dd y \,N(y) \,\mathcal{O}(x,y)}
         {\int_{S^1} \dd y\, N(y) }\,.
\end{equation}
For scalars this is the usual weighted average over the compact direction. When~$\mathcal{O}$ carries four-dimensional tensor indices, the
same notation denotes the component-wise zero-mode projection in the chosen ADM foliation. We will also make use of the~$y$-average of the 4d spacetime average,
\begin{equation} \label{weightedav2}
  \big\langle\! \langle \mathcal{O} \rangle \!\big\rangle_{S^1}
  \equiv \frac{\int_{S^1} \dd y\, N(y) \,\langle \mathcal{O} \rangle}
         {\int_{S^1} \dd y \,N(y) } = \frac{\int_{S^1} \dd y \,\frac{N(y)}{\mathcal{V}_4} \, \int_{\Sigma_y^\circ} \dd ^4x \sqrt{-g}\, \mathcal{O}(x,y)}
         {\int_{S^1} \dd y \,N(y) }  \,.
\end{equation}
This should not be confused with the average over the regulated five-dimensional region,
\begin{equation}
\langle \mathcal{O} \rangle_{\Sigma_y^\circ \times S^1}
  \equiv  \frac{
  \int_{S^1} \dd y\, N(y)
  \int_{\Sigma_y^\circ} \dd ^4x\,\sqrt{-g}\,
  \mathcal{O}(x,y)
  }{
  \int_{S^1} \dd y\, N(y)\,\mathcal{V}_{4}(y)
  }\, .
\end{equation}
Solving~\eqref{Neqn_review} for~$Q(y)$ and substituting back into~\eqref{geqn_review} gives the effective gravitational equations
\begin{equation}\label{geff_review}
    G_{\mu\nu} = \kappa^2 \left[ T^{(\text{m})}_{\mu\nu} - \frac{1}{2} \left(\big\langle T^{(\text{m})}\big\rangle - 2 \frac{\Gamma_{\rm m}}{\mathcal{V}_4} \right)  g_{\mu\nu} \right]\,.
\end{equation}
To see how vacuum energy decouples from these equations, let us split the matter effective action as
\begin{equation}
\Gamma_{\rm m}[g_y,\Psi_y] = - V_{\rm vac}\,\mathcal{V}_4(y) + \Gamma_{\rm dyn}[g_y,\Psi_y]\,,
\qquad
\Gamma_{\rm dyn}[g_y,\Psi_y^{\rm vac}]=0\, .
\label{Gam split z=0}
\end{equation}
Here $V_{\rm vac}$ is the renormalized coefficient of the local identity operator in the matter effective action, evaluated at the chosen matter
vacuum $\Psi_y^{\rm vac}$. Equivalently, it is the coefficient of the zero-derivative spacetime-volume term. The functional $\Gamma_{\rm dyn}$
contains the remaining (generally non-local) matter-dependent part of the effective action, normalized so that it vanishes when the matter fields are set to their vacuum expectation values.

The vacuum energy cancels identically from the gravitational field equations~\eqref{geff_review} and the dynamics reduce to
\begin{equation}
  G_{\mu\nu}
  = -\Lambda_{\text{eff}} g_{\mu\nu}
    + \kappa^2 T^{(\text{dyn})}_{\mu\nu}\,,
\label{Einstein eqns Lambda eff}
\end{equation}
with~$T^{(\text{dyn})}_{\mu\nu} = - \frac{2}{\sqrt{-g}} \frac{\delta \Gamma_{\rm dyn}}{\delta g^{\mu\nu}}$ and
\begin{equation}\label{Lambdaeffy}
    \Lambda_{\text{eff}}(y) = \frac{1}{2}\kappa^2\left[ \big\langle T^{(\text{dyn})}\big\rangle  -2 \frac{\Gamma_{\rm dyn}[g_y,\Psi_y]}{\mathcal{V}_4(y)} \right]\, .
\end{equation}
The central result is therefore the complete decoupling of the radiatively unstable vacuum energy generated by matter loops. This cancellation is algebraic: it occurs at fixed finite region $\Sigma_y^\circ$, before any regulator limit $r\to\infty$ is taken. Hence the decoupling of $V_{\rm vac}$ is independent of the IR regulator.
 
The residual quantity $\Lambda_{\rm eff}$ is by contrast a spacetime average of the dynamical, non-vacuum part of the matter effective action.
For non-compact $\Sigma_y$, its physical value is defined only when the regulated large-volume limit exists and is independent of how the region
$\Sigma_y^\circ$ is enlarged. Such a regulator-independent limit is expected in many physically reasonable situations. For example, in a universe
approaching a single asymptotic de Sitter phase, ordinary matter excitations are diluted and the regulated spacetime averages are governed by the
asymptotic state; localized or finite-duration excitations then do not contribute to the infinite-volume average. On the other hand, eternally
inflating geometries or strongly inhomogeneous multiverse-like configurations with competing asymptotic regions can make the regulated
average prescription-dependent. Such cases require an additional infrared measure or cutoff prescription and will not be considered here.

In contrast to~\cite{Carroll:2017gqo}, the decoupling of the matter-loop vacuum energy works to all orders in matter perturbation theory. This is
because the foliation-preserving diffeomorphisms protect the linear scaling of the vacuum-energy operator with respect to the projectable lapse, provided
the regulator and renormalization prescription preserve these symmetries.

Although the~$z=0$ theory is ultra-local in~$y$, with each slice evolving independently, for a small compact extra dimension, it is instructive to
project the equations onto their $y$-zero mode to obtain an alternative representation of the four-dimensional physics. Averaging the mixed-index components of~\eqref{Einstein eqns Lambda eff} over~$y$ gives
\begin{equation}
  \big\langle \tensor{G}{^\mu_\nu} \big\rangle_{S^1}
  = -\bar \Lambda_{\text{eff}} \delta^\mu_\nu
    + \kappa^2 \big\langle \tensor{T}{^{(\text{dyn})}^\mu_\nu} \big\rangle_{S^1}\,.
\end{equation}
This describes the averaged Einstein equations in terms of an average of the matter sources and an averaged effective cosmological constant,
\begin{equation}
  \bar \Lambda_{\text{eff}}= \big\langle \Lambda_\text{eff} \big\rangle_{S^1}
  = \frac{1}{2} \kappa^2\left\langle \big\langle T^{(\text{dyn})}\big\rangle  -2 \frac{\Gamma_{\rm dyn}[g_y,\Psi_y]}{\mathcal{V}_4(y)}\right\rangle_{S^1}\,.
\end{equation}
Here~$\langle \tensor{G}{^\mu_\nu}\rangle_{S^1}$ is the zero-mode projection of the Einstein tensor computed from~$g_{\mu\nu}(x,y)$ on each slice; it is
not, in general, the Einstein tensor of an averaged metric. Of course, radiative corrections to the vacuum energy are absent from this averaged equation, just as they are absent from \eqref{Lambdaeffy}. This additional averaging is not essential in the~$z=0$ limit, but it will become especially important in the next section when we deform the theory away from ultra-locality and allow neighbouring slices to interact.

We end this section with a comment on early universe phase transitions, where the vacuum energy changes over a short period of time from one constant value to another. These were discussed in detail in the context of vacuum energy sequestering, where it was shown that the effects are diluted by the large age of the universe \cite{Kaloper:2014dqa}. The same is true here. To see this, imagine a universe that undergoes a single phase transition. We assume that the phase transition completes itself (no eternal inflation), and does so rapidly on a Hubble time, such that it can be approximated as occurring instantaneously. The corresponding spacelike surface~$\Sigma_\text{PT}$ splits the spacetime manifold into two regions~$\mathcal{M}_-$ (before the transition) and~$\mathcal{M}_+$ (after). Neglecting the localized stress-energy of the transition surface, the energy-momentum tensor is given by~$T_{\mu\nu}=-V g_{\mu\nu}$, with~$V=V_\pm$ in~$\mathcal{M}_\pm$. 

A constant shift of~$V$ cancels from the Einstein equations~\eqref{geff_review}; only the spacetime variation associated with the transition contributes. Indeed, 
we have that~$G_{\mu\nu}=-\kappa^2 V_\text{eff} g_{\mu\nu}$, where
\begin{equation}
  V_\text{eff}=  V-\langle V\rangle = \begin{cases}
      -\Delta V (1- \epsilon)  & \text{in~$\mathcal{M}_-$}\\
      \Delta V \epsilon  & \text{in~$\mathcal{M}_+$}
  \end{cases}
\end{equation}
where~$\Delta V=V_+-V_-$ is the jump in vacuum energy, and 
\begin{equation}
\epsilon=\frac{\int_{\mathcal{M_-}} \dd ^4 x \sqrt{-g} }{{\int_{\mathcal{M_-} \cup \mathcal{M_+}} \dd ^4 x \sqrt{-g}}}
\end{equation}
is the regulated fractional volume of the universe prior to the transition. For an early phase transition in a universe whose later evolution approaches a single asymptotic de Sitter phase, the post-transition four-volume dominates the regulated spacetime average. In this case~$\epsilon$ becomes extremely small as the infrared region is enlarged, and the late-time curvature sourced by the transition is correspondingly diluted.\footnote{Localized excitations, finite-duration non-adiabatic effects, and the stress-energy of the transition surface are suppressed for the same reason, provided their total regulated contribution grows more slowly than the spacetime volume.} This conclusion assumes that the regulated spacetime average has a unique large-volume limit. Eternally inflating geometries, or strongly inhomogeneous multiverse-like configurations with competing asymptotic regions, can make the fractional volumes prescription-dependent. Such cases require an additional infrared measure or cutoff prescription and are not addressed here.

\section{Bulk dynamics for gravity and fluxes} \label{sec:def}

We now relax the ultra-local approximation only in the classical gravitational and higher-form sectors. We include the leading operators
containing derivatives normal to the foliation, at the same derivative order as the intrinsic four-dimensional kinetic terms under a~$z=1$
counting, while retaining an ultra-local matter sector. This is a semiclassical, or more precisely matter-loop, truncation: the
metric and form fields are treated as classical backgrounds, whereas quantum corrections are computed only from matter fluctuations. Because
the microscopic matter action contains neither normal derivatives nor interactions coupling distinct~$y$-slices, the matter functional integral factorises slice by slice. Graviton or higher-form loops, or the inclusion of normal-derivative matter operators, would in
general invalidate this truncation and are not included here. 

The normal-derivative terms retained below should therefore be regarded as independent classical bulk EFT operators. They
couple neighbouring slices in the gravitational and flux sectors and become important at shorter wavelengths along the compact direction.
For special choices of their coefficients they assemble into five-dimensional Lorentz-invariant kinetic combinations. 

The covariant way to introduce $ y $-derivatives of the spacetime metric $ g_{\mu\nu} $ is through the extrinsic curvature of constant-$ y $ slices, defined as
\begin{equation}
    H_{\mu\nu} = \frac{1}{2}\mathcal{L}_n  g_{\mu\nu} \,,
\end{equation}
where~$\mathcal{L}_n\equiv\frac{1}{N} \big(\del_y-\mathcal{L}_{N^\mu}\big)$, with~$ \mathcal{L}_{N^\mu} $ denoting the Lie derivative along the shift vector~$ N^\mu $.
The extrinsic curvature is the appropriate building block because it transforms covariantly under foliation-preserving diffeomorphisms.
For~$ z = 1 $, the gravitational part of the action takes the form
\begin{equation}
    S_\text{grav} = \frac{1}{2\kappa^2} \int \dd y\, N \int_{\Sigma_y} \dd^4x\, \sqrt{-g} \, \Big( R^{(4)}- \lambda H_{\mu\nu} H^{\mu\nu} +\mu H^2 \Big)+\text{boundary terms} \,,
\label{z=1 gravitational action}
\end{equation}
where $ H = g^{\mu\nu} H_{\mu\nu} $. The choice $ \lambda = \mu = 1 $ recovers the Lorentz-invariant combination familiar from General Relativity. For~$\lambda,\mu \neq 1$, we can fix either~$\lambda$ or~$\mu$ to unity (but not both) by an appropriate rescaling of the lapse. The boundary contributions correspond to the Gibbons--Hawking--York term evaluated on the boundary $ \partial \Sigma_y $ of each constant-$ y $ slice, as discussed previously. Owing to the compactness of the extra dimension and the imposition of periodic boundary conditions, the five-dimensional manifold is without boundary in the $ y $-direction. As such, no further boundary terms arise from the integration over $ 
 y$.
  
We now consider the three-form field $A_3$. This can either be incorporated into an $n$-form~$\mathcal{B}_n$ ($n=3,4$) in five dimensions in two distinct ways. Either it is promoted to a five-dimensional three-form field as
\begin{equation}
  \mathcal{B}_3(x, y) = A_3(x, y) + N \, \dd y \wedge A_2(x, y)\,,
\end{equation}
or, alternatively, to a five-dimensional four-form potential as
\begin{equation}
  \mathcal{B}_4(x, y) = A_4(x, y) + N \, \dd y \wedge A_3(x, y)\,.
\label{B4}
\end{equation}
Generalising Eq.~\eqref{eq:A3definition}, we define
\begin{equation}
  A_n(x, y) \equiv \frac{1}{n!}\, A_{\mu_1 \ldots \mu_n}(x, y) \,\theta^{\mu_1} \wedge \dots \wedge \theta^{\mu_n},
\end{equation}
where we recall that the co-frames are~$\theta^{\mu} = \dd x^{\mu} + N^{\mu}(x, y) \, \dd y$. In other words,~$A_3$ can either be built into a 5d three-form~$\mathcal{B}_3$ by introducing a two-form~$A_2$, or
to a 5d four-form~$\mathcal{B}_4$ by introducing a four-form~$A_4$. From the~$n$-form~$\mathcal{B}_n$ ($n=3,4$) we can define a corresponding~$(n+1)$-form field strength in five dimensions, where
\begin{equation}
    \mathcal{F}_{n+1}\equiv \dd_5 \mathcal{B}_n= F_{n+1}+N\, \dd y \wedge (C_n-F_{n}) \,,
\end{equation}
with~$F_{n}=\dd_4 A_{n-1}$, and
\begin{equation}
    C_n=\frac{1}{n!}  \mathcal{L}_n A_{\mu_1 \ldots \mu_n }\dd x^{\mu_1} \wedge \ldots \wedge \dd x^{\mu_n} \,.
\end{equation}
 
While both cases are in principle allowed, it is straightforward to see that the case~$n = 3$ can be dangerous phenomenologically. Indeed, for $n = 3$ the most general $z = 1$ flux operators are constructed from the four-form $F_4$, its dual $\star_4 F_4$, as well as from the three-forms $F_3$ and $C_3$, along with their respective duals. If the three-forms acquire non-trivial vacuum expectation values, they will select a preferred direction in the four-dimensional spacetime, thereby breaking 4d Lorentz invariance. This issue does not arise for $n = 4$, where the $z = 1$ operators are built exclusively from the four-forms $F_4$ and $C_4$, and their duals. For this reason, we henceforth focus on the case $n = 4$, where our spacetime three-form field $A_3$ is embedded into a four-form~$\mathcal{B}_4$ in five dimensions given by Eq.~\eqref{B4}. 

For~$z=1$,~$A_3$ thereby enters the action as follows
 \begin{equation}
    S_\text{$p$-forms}=\frac{1}{2\kappa^2} \int \dd y\, N \int_{\Sigma_y} \bigg( -F_4 \wedge \star_4 F_4 -2\alpha F_4 \wedge \star_4 C_4   - \beta C_4\wedge \star_4 C_4 \bigg) 
    +\text{boundary terms} \,,
\label{z=1 flux action}
\end{equation}
where~$\alpha,\beta$ are constants. If~$\alpha \neq 0$, we can set it to unity by absorbing it into a redefinition of~$A_4$.\footnote{Explicitly, 
define~$\widetilde A_4=\alpha A_4$, and hence~$\widetilde C_4=\alpha C_4$. This sets the coefficient of the mixed term to unity while replacing~$\beta$ by $\widetilde\beta=\beta/\alpha^2$. Since we retain~$\alpha$ explicitly in what follows, we will not make this redefinition.}
The Lorentz-invariant limit corresponds to the case where~$\beta=\alpha^2$, in which case this is just the canonical kinetic term for the five-form field strength in five dimensions. The boundary terms generalise the Duncan-Jensen term~\eqref{DJ_review} encountered earlier. For Neumann boundary conditions, they are given by 
\begin{equation} \label{DJz=1}
    \frac{1}{\kappa^2} \int \dd y\, N \int_{\partial \Sigma_y} A_3 \wedge \star_4 \big(F_4+\alpha C_4\big)\,.
\end{equation}
  Standard Neumann boundary conditions on~$A_3$ correspond to~$\delta \left[ \star_4 \big(F_4+\alpha C_4\big) \right]=0$ on the boundary~$\partial\Sigma_y$. Since we also have Neumann boundary conditions on the lapse, we can generalise the Neumann boundary condition on~$A_3$ to be of the form
\begin{equation} \label{deltaF}
    \delta  \Big[ \star_4 \big(F_4+\alpha C_4\big)\Big] =  \alpha\big({\star_4 C_4}\big)\frac{\delta N}{N} \,.
\end{equation}
As we will see in a moment, the form of this boundary condition is chosen to recover the correct form for the global constraint arising from the variation with respect to the lapse. The modified Neumann-like boundary condition~\eqref{deltaF} is analogous to the modified Dirichlet-like boundary condition~\eqref{bc_review} for the metric.

Combining the gravitational part~\eqref{z=1 gravitational action} and the flux part~\eqref{z=1 flux action} of the action, together with matter effective action, 
we obtain the full theory:
\begin{equation} \label{action-decomp}
\begin{split}
   S = \frac{1}{2\kappa^2} &\int \dd y\, N \int \dd^4x \bigg[ \sqrt{-g}  \Big(R^{(4)}-\lambda H_{\mu\nu} H^{\mu\nu}+\mu H^2\Big)
 - F_4 \wedge \star_4   F_4 -2\alpha  F_4 \wedge \star_4   C_4-\beta C_4\wedge \star_4 C_4 \bigg]
 \\ 
+ &\int \dd y\, N\, \Gamma_{\rm m}[g_y,\Psi_y]  + \text{boundary terms} \,.
\end{split}
\end{equation}
As before, the matter effective action is assumed to be ultra-local in~$y$. The functional~$\Gamma_{\rm m}[g_y,\Psi_y]$ is intrinsic to each slice: it contains no dependence on the shift, or normal derivatives of the matter fields. A renormalization prescription that preserves this factorisation and the foliation-preserving diffeomorphisms therefore generates intrinsic four-dimensional functionals on each slice, but no matter propagation or non-locality in the~$y$-direction.

\subsection{Equations of motion}

The field equations from varying~$A_3$ imply that
\begin{equation}
\star_4 \big(F_4+\alpha C_4\big)=Q(y)\,, 
\end{equation}
where~$Q$ is a function of~$y$ only. From the variation of~$A_4$ we obtain
\begin{equation}
    \mathcal{L}_n\Big[{\star_4 (\alpha F_4+\beta C_4)}\Big]=0\,.
\label{A4 eom}
\end{equation}
For simplicity, we will set the shift vector to vanish on our solutions,
\begin{equation}
    N^\mu = 0\,.
\end{equation}
In this case the solution to Eq.~\eqref{A4 eom} is readily obtained,
\begin{equation}
\star_4 \big(\alpha F_4+\beta C_4\big)=P(x)\,.
\end{equation}
Here~$P$ is a function of~$x^\mu$ only. For~$\beta=\alpha^2$, we necessarily have~$P(x)=\alpha Q(y)=$ constant. For~$\beta \neq \alpha^2$, we find
\begin{equation}
\begin{split}
    \star_4 F_4&=\frac{\alpha P(x)-\beta Q(y)}{\alpha^2-\beta}\,,\\ 
    \star_4 C_4 &=\frac{-P(x)+\alpha Q(y)}{\alpha^2-\beta}\,.
    \label{star F4 C4}
\end{split}
\end{equation}
In the rest of the section we will focus on this latter case ($\beta \neq \alpha^2$), leaving the Lorentz-invariant couplings~($\beta = \alpha^2$ and~$\mu = \lambda$) to Sec.~\ref{sec:stuck}. 

The equation of motion from the lapse receives non-trivial contributions from the boundary conditions for the three-form~\eqref{deltaF} and the metric~\eqref{bc_review}, giving rise to the following constraint,
\begin{equation} \label{Neq}
    0=\int \dd ^4 x \sqrt{-g} \left[ \frac{1}{2}\big(R^{(4)} +\lambda H_{\mu\nu}H^{\mu\nu}-\mu H^2\big) \right. + \left.\frac{1}{2(\alpha^2-\beta)} \mathcal{Q}   \right] +\kappa^2 \Gamma_{\rm m}[g_y,\Psi_y] \,.
\end{equation}
Here we have defined the total flux contribution
\begin{equation}
\begin{split}
    \mathcal{Q} &= P^2 -2\alpha P Q+\beta Q^2 \\
    &= \left(P-\alpha Q\right)^2+\big(\beta-\alpha^2\big)Q^2\,.
\label{F def}
\end{split}
\end{equation}
In contrast, the equation of motion from the variation of the shift vector gives an equation that is local on the spacetime slices, 
\begin{equation} \label{Nnu}
   \nabla^\mu \big( \lambda H_{\mu\nu} -\mu H g_{\mu\nu}\big) =\star_4 A_4 \nabla_\nu P\,.
\end{equation}
This can be understood as an evolution equation for the flux~$P(x)$. Similarly, the Einstein equations are also local and given by
\begin{equation}  \label{Ein}
G_{\mu\nu}=\frac{1}{2(\alpha^2-\beta)}\mathcal{Q} g_{\mu\nu}+\kappa^2\left(T^{(H)}_{\mu\nu}+T^{(\text{m})}_{\mu\nu}\right)\,,
\end{equation}
where, as before,~$T^{(\text{m})}_{\mu\nu}=-\frac{2}{\sqrt{-g}}\frac{\delta \Gamma_{\rm m}}{\delta g^{\mu\nu}}$ is the energy--momentum tensor for the matter sector.
The contribution from the extrinsic curvature terms is given by
\begin{equation}
\begin{split}
T^{(H)}_{\mu\nu} &\equiv -\frac{2}{N\sqrt{-g}}\frac{\delta }{\delta g^{\mu\nu}} \int \dd y\, N(y)  \int \dd ^4 x \sqrt{-g}\frac{1}{2\kappa^2} \Big(-\lambda H_{\mu\nu}H^{\mu\nu} +\mu H^2\Big)   \\
 &= \frac{1}{2 \kappa^2} \bigg(-4\lambda  \tensor{H}{_\mu^\alpha} H_{\alpha \nu}+2(\lambda+2 \mu) HH_{\mu\nu}  -g_{\mu\nu }\left(\lambda H_{\alpha \beta}H^{\alpha\beta} +\mu H^2\right) + 2\mathcal{L}_n (\lambda H_{\mu\nu}-\mu H g_{\mu\nu})\bigg)\,.
\end{split}
\label{TH 0}
\end{equation}
Later it will be convenient to express this with one index raised, 
\begin{equation} \label{TH}
  \tensor{T}{^{(H)\mu} _\nu}  =\frac{1}{2 \kappa^2} \bigg(2\lambda H\tensor{H}{^\mu_\nu}  -\delta^\mu_{\nu}\left(\lambda H_{\alpha \beta}H^{\alpha\beta} +\mu H^2\right)    + 2\mathcal{L}_{n} \big(\lambda \tensor{H}{^\mu_\nu}-\mu H \delta^\mu_{\nu} \big)\bigg)\,.
\end{equation}

Note that~$P$ and~$Q$, which parametrize the flux contribution, appear in the same combination~$\mathcal{Q} = P^2 -2\alpha P Q+\beta Q^2$ in both the lapse global constraint~\eqref{Neq} and the metric equations~\eqref{Ein}. This is a direct consequence of the careful choice of boundary conditions and is key to achieving the desired cancellation of vacuum energy. As we are about to show, that cancellation is carried out by the global part of the total flux contribution~$\left\langle \mathcal{Q} \right\rangle$. To see this, we divide the lapse equation~\eqref{Neq} by the 4d regulated spacetime volume~$\mathcal{V}_4(y)=\int_{\Sigma_y^\circ} \dd^4 x\,\sqrt{-g}$, introduced in Eq.~\eqref{volume def}, such that it becomes a constraint on the average 4d scalar curvature,
\begin{equation}
    \big\langle R^{(4)} \big\rangle = -\frac{1}{\alpha^2 - \beta} \left\langle \mathcal{Q} \right\rangle - \left\langle \lambda  H_{\alpha\beta} H^{\alpha\beta}  - \mu H^2 \right\rangle - \frac{2\kappa^2}{\mathcal{V}_4(y)} \Gamma_{\rm m} \, .
\end{equation}
Similarly, taking the trace and spacetime average of the metric equation~\eqref{Ein} yields a second global constraint on the curvature,
\begin{equation}
    \big\langle R^{(4)} \big\rangle = -\frac{2}{\alpha^2 - \beta} \left\langle \mathcal{Q} \right\rangle  - \kappa^2 \left(\big\langle T^{(H)} \big\rangle + \big\langle T^{\text{(m)}} \big\rangle \right) \, .
\end{equation}
Eliminating~$ \langle R^{(4)} \rangle$ from these two constraints fixes the average of the flux contributions to be
\begin{equation} \label{constr}
   0 = \frac{1}{2(\alpha^2 - \beta)}
    \left\langle \mathcal{Q} \right\rangle  
    - \frac{1}{2}\left\langle  \lambda H_{\alpha\beta} H^{\alpha\beta} -\mu H^2 \right\rangle +\frac{\kappa^2}{2} \left(\big\langle T^{(H)} \big\rangle
    + \big\langle T^{\text{(m)}} \big\rangle\right) - \frac{\kappa^2}{\mathcal{V}_4(y)} \Gamma_{\rm m} \,.
\end{equation}
One should think of this equation as fixing the average flux contribution in terms of the overall cosmological constant. Combining Eqs.~\eqref{Ein} and~\eqref{constr} yields effective Einstein equations of the form
\begin{equation}  \label{Ein2}
\begin{split}
\tensor{G}{^\mu_\nu}=\frac{1}{2(\alpha^2-\beta)}\Delta \mathcal{Q} \delta^\mu_\nu
&+\kappa^2 \left(\tensor{T}{^{(H)\mu}_\nu} -\frac{1}{2} \big\langle T^{(H)} \big\rangle  \delta^\mu_\nu\right) +\frac{1}{2} \big\langle \lambda H_{\alpha\beta} H^{\alpha\beta} -\mu H^2 \big\rangle  \delta^\mu_\nu \\
&+\kappa^2 \left(\tensor{T}{^{(\text{m})}^\mu_\nu}-\frac{1}{2} \big\langle T^{\text{(m)}} \big\rangle  \delta^\mu_\nu\right) +\frac{\kappa^2}{\mathcal{V}_4(y)} \Gamma_{\rm m}  \delta^\mu_\nu  \,,
\end{split}
\end{equation}
where~$\Delta\mathcal{Q}=\mathcal{Q}  - \langle \mathcal{Q} \rangle$ is the local fluctuation in the flux.

For the~$z=1$ theory, the~$y$-slices do not evolve independently. Recall that the extra dimension is assumed to be small and compact. To get the four-dimensional perspective, it is therefore natural to take an average over the extra dimension, as per Eqs.~\eqref{weightedav} and~\eqref{weightedav2}. To do this, we first note that 
\begin{equation}\label{eq:THmunu_average}
 \kappa^2 \Big\langle \tensor{T}{^{(H)}^\mu_\nu} \Big\rangle_{S^1}=\Big\langle \lambda H\tensor{H}{^\mu_\nu}  -\frac{1}{2}\delta^\mu_\nu\Big(\lambda H_{\alpha \beta}H^{\alpha\beta} +\mu H^2\Big) \Big\rangle_{S^1}\,,  
\end{equation}
where the contribution from~$\mathcal{L}_{n} \big(\lambda \tensor{H}{^\mu_\nu} -\mu H \delta^\mu_\nu\big)$ has dropped out after integration over the extra dimension, on account of the periodic boundary conditions, and setting~$N^\mu = 0$. To evaluate~$\kappa^2 \left\langle\! \big\langle T^{(H)} \big\rangle\! \right\rangle_{S^1}$, we make use of the fact that, for any scalar function~$\varphi(x,y)$,
\begin{equation}
    \int_{S^1} \dd y\, N(y) \langle \mathcal{L}_n \varphi \rangle=\int_{\Sigma_y^\circ \times S^1} \dd y\, \dd^4 x\, N(y)  \mathcal{L}_n\left(\frac{\sqrt{-g}  \varphi }{{\cal V}_4(y)}\right)+\int_{S^1} \dd y\, N(y) \Big( \langle \varphi \rangle \langle H \rangle -\langle \varphi H \rangle\Big)\,,
\label{Lnphi identity}
\end{equation}
where we have assumed, for simplicity, that the finite infrared region on each slice,~$\Sigma_y^\circ\subset\Sigma_y$, is independent of~$y$. That is, the regulating region is homogeneous in~$y$. Using~\eqref{Lnphi identity}, we obtain
\begin{equation}
  \kappa^2\Big\langle\! \big\langle T^{(H)} \big\rangle \!\Big\rangle_{S^1}=\Big\langle  -2 \lambda \big\langle H_{\alpha \beta} H^{\alpha \beta}\big\rangle+2 \mu \big\langle H^2  \big\rangle+(\lambda-4 \mu) \big\langle H\big\rangle^2 \Big \rangle_{S^1} \,.
\end{equation}
Once again, the contribution from a total derivative along the extra dimension has dropped out on account of the periodic boundary conditions. Putting everything together, the four-dimensional Einstein equations averaged over the extra dimension are found to be
\begin{equation}  \label{Ein3}
\begin{split}
\big\langle \tensor{G}{^\mu_\nu} \big\rangle_{S^1} = \bigg\langle &\frac{1}{2\big(\alpha^2-\beta\big)} \Delta\mathcal{Q}\, \delta^\mu_\nu\\
&+\frac{\lambda}{2}{} \left(
 2H\tensor{\widehat{H}}{^\mu_\nu}- \left( \widehat H_{\alpha \beta}\widehat{H}^{\alpha \beta}-3\big\langle \widehat{H}_{\alpha \beta}\widehat{H}^{\alpha \beta} \big\rangle \right)\delta^\mu_\nu \right)   +\frac{\lambda-4 \mu}{8} \left( H^2 +3\big\langle H^2 \big\rangle -4 \big\langle H\big\rangle^2\right) \delta^\mu_\nu\\
&+\kappa^2 \left( \tensor{T}{^{(\text{m})}^\mu_\nu}-\frac{1}{2} \langle T^{\text{(m)}} \rangle \delta^\mu_\nu
+ \frac{\Gamma_{\rm m}}{\mathcal{V}_4(y)} \delta^\mu_\nu \right)\bigg\rangle_{S^1} \,,
\end{split}
\end{equation}
where~$\tensor{\widehat{H}}{^\mu_\nu}=\tensor{H}{^\mu_\nu}-\frac{1}{4} H\delta^\mu_\nu$ is the traceless part of the extrinsic curvature. 

\subsection{Vacuum energy cancellation}
To understand how vacuum energy enters the effective gravitational equations, we split the matter effective action, as we did in Eq.~\eqref{Gam split z=0}, to extract the vacuum energy contribution.
\begin{equation}
    \Gamma_{\rm m}[g_y,\Psi_y] = - V_{\rm vac}\,\mathcal{V}_4(y) + \Gamma_{\rm dyn}[g_y,\Psi_y]
\label{Gam split z=1}
\end{equation}
As before,~$\Gamma_{\rm dyn}$ encodes the generally non-local, matter-dependent part of the effective action, which vanishes on the matter vacuum,~$\Gamma_{\rm dyn}[g_y,\Psi_y^{\rm vac}]=0$. Matter-induced geometric terms, such as local curvature counterterms, are assigned to the renormalized gravitational action.\footnote{Assuming ultra-locality is essential in this decomposition. Had we included normal-derivative operators in the matter sector, the matter loop determinants would involve the full Kaluza--Klein spectrum, and the total matter effective action would then acquire a compactification-dependent vacuum functional~$\Gamma_{\rm comp}^{\rm vac}[g,N;L]$, where~$L=\int_{S^1}\dd y\,N(y)$ is the proper length of the compact dimension. Its finite part would contain Casimir-like terms with non-trivial dependence on~$L$. The ultraviolet part would renormalize local bulk geometric operators, whose integrated contributions remain extensive in~$L$. Such contributions modify the lapse constraint and need not obey the cancellation exhibited below for an ultra-local vacuum energy. They will be considered in more detail in Sec.~\ref{Casimir sec}.}

Substituting the matter stress tensor~$T^\text{(m)}_{\mu\nu} =-V_\text{vac} g_{\mu\nu}+T^\text{(dyn)}_{\mu\nu}$ into Eq.~\eqref{Ein3},
we see that the vacuum energy drops out. Indeed,
\begin{equation}  \label{Ein4}
\begin{split}
\langle \tensor{G}{^\mu_\nu} \rangle_{S^1} = \bigg\langle &\frac{1}{2\big(\alpha^2-\beta\big)}  \Delta\mathcal{Q}\, \delta^\mu_\nu\\
&+\frac{\lambda}{2}{} \left(
 2H\tensor{\widehat{H}}{^\mu_\nu}- \left( \widehat H_{\alpha \beta}\widehat H^{\alpha \beta}-3\big\langle \widehat H_{\alpha \beta}\widehat H^{\alpha \beta} \big\rangle \right)\delta^\mu_\nu \right)   +\frac{\lambda-4 \mu}{8} \left( H^2 +3\big\langle H^2 \big\rangle -4 \big\langle H\big\rangle^2\right) \delta^\mu_\nu\\
&+ \kappa^2 \left( \tensor{T}{^{(\text{dyn})}^\mu_\nu}-\frac{1}{2} \big\langle T^{\text{(dyn)}} \big\rangle \delta^\mu_\nu  + \frac{\Gamma_{\rm dyn}}{\mathcal{V}_4(y)}
  \delta^\mu_\nu\right) \bigg\rangle_{S^1} \,.
\end{split}
\end{equation}
The last line contains only dynamical, non-vacuum matter contributions.

Of course, we might worry that the vacuum energy feeds into the curvature indirectly through the extrinsic curvature terms (second line in Eq.~\eqref{Ein4}). To see that this is not  a problem, let us focus on the vacuum energy contribution by setting~$\Gamma_{\rm dyn} = 0$ and~$\tensor{T}{^{(\text{dyn})}^\mu_\nu} = 0$. We also assume that the field ansatz preserves maximal symmetry along the~$y$–slicing. Accordingly, the four-dimensional metric takes the warped form
\begin{equation}
   g_{\mu\nu} = a^2(y)\, q_{\mu\nu}(x)\,,
\label{warped}
\end{equation}
where~$q_{\mu\nu}(x)$ is a maximally symmetric metric with constant Ricci scalar, {\it i.e.},~$R^{(4)}(q) = {\rm const}$. Recall that the lapse function depends only on~$y$,~$N = N(y)$, and the shift vector vanishes by assumption,~$N^\mu = 0$. Choosing the gauge~$N(y) = 1$, it follows that
\begin{equation} \label{extH}
\tensor{H}{^\mu_\nu} =\frac{a'}{a} \delta^\mu_\nu\,.
\end{equation}
Maximal symmetry further requires that both fluxes be constant on each slice~$\Sigma_y$. 
This condition is automatically satisfied for~$Q(y)$, but it constrains~$P(x)$ to be constant.

With this ansatz, the lapse equation \eqref{Neq} gives 
\begin{equation} \label{Nvac}
     -\frac{R^{(4)}(q)}{2a^2} = \frac{1}{2(\alpha^2-\beta)} \mathcal{Q} +2(\lambda -4 \mu) \left( \frac{a'}{a}\right)^2   
  -\kappa^2 V_\text{vac} \,.
\end{equation}
The shift equation \eqref{Nnu} is satisfied automatically while the metric equation \eqref{Ein} gives
\begin{equation}  \label{Einvac}
-\frac{R^{(4)}(q)}{4a^2}=\frac{1}{2(\alpha^2-\beta)} \mathcal{Q}  +(\lambda-4\mu)\left(2\left( \frac{a'}{a}\right)^2 +\left( \frac{a'}{a}\right)' \right)-\kappa^2 V_\text{vac}\,,
\end{equation}
Eliminating the flux contributions from Eqs.~\eqref{Nvac} and \eqref{Einvac} yields
\begin{equation}  \label{Einvac no flux}
\frac{R^{(4)}(q)}{4a^2}=(\lambda-4\mu)\left( \frac{a'}{a}\right)'\,.
\end{equation}
The key feature here is the fact that the right hand side is a total derivative in $y$. 

Integrating over the compact extra dimension and invoking periodic boundary conditions,  it follows that
\begin{equation}
R^{(4)}(q)=0\,.
\end{equation}
In other words, the curvature of the four-dimensional vacuum must vanish, regardless of the vacuum energy. We can also solve directly for the warp factor. Assuming $\lambda \neq 4\mu$, we find that~$a(y) =c_1 e^{c_2y}$ for constants~$c_1, c_2$. Applying periodic boundary conditions we find that~$c_2=0$, and so the warp factor is a constant:
\begin{equation}
a(y) = {\rm constant}\,.
\end{equation}
The special case~$\lambda=4\mu$ corresponds to a extrinsic-curvature conformal point: 
the extrinsic-curvature quadratic form reduces to~$-\lambda H_{\mu\nu}H^{\mu\nu}+\mu H^2
=-4\mu\,\widehat H_{\mu\nu}\widehat H^{\mu\nu}$, which is therefore insensitive to the trace, or conformal, part of the
extrinsic curvature. For this special conformal point, the warp factor~$a(y)$ is arbitrary.\footnote{This is the four-dimensional version of the critical value in conformal Hořava-type kinetic terms~\cite{Bellorin:2023nuh}. Although this kinetic sector admits an anisotropic Weyl invariance when the lapse is assigned conformal weight four, the full~$z=1$ action considered here is not Weyl invariant.} Lastly, from either Eq.~\eqref{Nvac} or~\eqref{Einvac}, the solution for the background flux is
\begin{equation}
\mathcal{Q} = 2\kappa^2 \big(\alpha^2-\beta\big) V_{\rm vac}\,.
\label{F vac}
\end{equation}
The existence of a real flux solution imposes a restriction on the parameter branch.\footnote{\label{Damienfootnote}We thank Damien Easson for pointing this out.} Indeed, if~$\beta>\alpha^2$,
we see from Eq.~\eqref{F def} that~$\mathcal{Q}$ is non-negative, which is consistent with Eq.~\eqref{F vac} only for~$V_{\rm vac}\leq0$ in our sign conventions.
If~$\beta < \alpha^2$, on the other hand, the solution exists for~$V_{\rm vac}$ of either sign. Thus, on the branch admitting a real flux solution, the four-dimensional vacuum slices are exactly flat because the flux contribution is forced to cancel the vacuum energy in the global constraint, together with periodic boundary conditions.

Equivalently, we can derive the flatness of the spacetime slices directly from Eq.~\eqref{Ein4}. Using the form of the extrinsic curvature~\eqref{extH}, we find that~$\tensor{\widehat{H}}{^\mu_\nu}=0$. Moreover, since~$H=4a'/a$ depends only on~$y$, it is constant over each slice~$\Sigma_y$, and hence~$\langle H\rangle=H$ and~$\langle H^2\rangle=H^2$. Consequently,~$H^2+3\langle H^2\rangle-4\langle H\rangle^2=0$, and all of the extrinsic-curvature terms in~\eqref{Ein4} vanish. Since both $P$ and $Q$ are constant on each slice~$\Sigma_y$, it follows that~$\Delta\mathcal{Q}=0$. The effective gravitational equations now give the following equation for the Ricci scalar
\begin{equation}
  R^{(4)}(q) \left \langle \frac{1}{a^2}\right\rangle_{S^1} =0 \qquad \implies \qquad R^{(4)}(q)  = 0  \,.
\end{equation}
Note that in deriving Eq.~\eqref{Ein4}, gradients of extrinsic curvature were eliminated after averaging over the extra dimension and making use of the periodic boundary conditions. This was crucial in ensuring the vanishing curvature in the vacuum. 

In summary, for the~$z=1$ theory, the interplay between the flux sector, the lapse constraint, and the choice of boundary conditions ensures that vacuum energy does not gravitate. The global flux constraint~\eqref{constr} fixes the net contribution of the higher-form fields, while the periodic geometry of the compact dimension eliminates any residual curvature. As a result, the four-dimensional spacetime remains exactly flat in vacuum, realising a dynamical cancellation of the cosmological constant. Thus the~$z=1$ deformation preserves vacuum-energy cancellation at the level of the background equations. Whether the resulting bulk dynamics is perturbatively consistent is a separate question, to which we now turn.

\section{Linearised perturbations and massive gravitons} \label{sec:lin}

In this section we study perturbations about the flat vacuum configuration of the~$z=1$ theory. More precisely, we expand about the sequestered vacuum found in the previous section: the four-dimensional metric is flat, the warp factor is constant, and the flux sector has adjusted so that the net four-dimensional curvature vanishes, even in the presence of a constant matter vacuum energy. After a constant rescaling of coordinates, we take
\begin{equation}
g_{\mu\nu} = \eta_{\mu\nu} + h_{\mu\nu}(x,y)\,, \qquad N = 1+\varphi(y)\,, \qquad
N^\mu = n^\mu(x,y)\,.
\label{linear perturbations}
\end{equation}
The background four-form fluxes need not vanish when~$V_{\rm vac}\neq0$. Nevertheless, the higher-form sector carries no local propagating degrees of freedom. Its principal role in the vacuum is to enforce the global constraint that cancels the cosmological-constant source. As we will show explicitly, flux perturbations affect only the spacetime-averaged sector and do not modify the local non-Fierz–Pauli structure of the spin-two quadratic action.

Expanding the action~\eqref{action-decomp} to quadratic order, it is useful to separate the local quadratic action from the term involving the projectable lapse fluctuation.
\begin{equation}
\delta_2S = \delta_2S_{\rm local} + \frac{1}{2\kappa^2} \int\dd y\,\varphi(y)\, {\cal C}^{(1)}(y) +
\text{boundary terms}
\label{quadratic action decomposition}
\end{equation}
Here~${\cal C}^{(1)}(y)$ denotes the first variation of the lapse constraint on the slice~$\Sigma_y$. The second term in Eq.~\eqref{quadratic action decomposition} is not an additional independent contribution to the action. Rather, it collects all terms proportional to~$\varphi(y)$ that arise from the quadratic expansion of the gravitational, flux, and matter sectors. Since the lapse is projectable,~$\varphi$ depends only on~$y$ and therefore imposes one spacetime-integrated constraint on each four-dimensional slice.

The local gravitational contribution is
\begin{equation}
\begin{split}
\delta_2S^{\rm grav}_{\rm local} = \frac{1}{2\kappa^2} \int\dd y \int_{\Sigma_y}\dd^4x\,
\bigg[ \delta_2\!\left(\sqrt{-g}\,R^{(4)}\right) - \lambda\delta_1H_{\mu\nu}\, \delta_1H^{\mu\nu}
+ \mu\big(\delta_1H\big)^2 \bigg]\,,
\end{split}
\label{quadratic gravitational action}
\end{equation}
where indices are raised and lowered with the background Minkowski metric.

The treatment of the higher-form sector requires some care because the background field strengths are generally non-zero. Although the original action is quadratic in~$F_4$ and~$C_4$, its expansion about a non-zero flux background contains linear and quadratic fluctuation terms, as well as couplings to the metric and lapse perturbations. It is convenient to first solve the higher-form equations in terms of the integration functions~$P(x)$ and~$Q(y)$, including the Duncan–Jensen boundary term, and then expand the resulting reduced action.

Recall the total flux combination~${\cal Q} = P^2 - 2\alpha P Q + \beta Q^2$ defined in Eq.~\eqref{F def}.
We write
\begin{equation}
\begin{split}
P &= \bar P+\delta P\,,\\
Q &= \bar Q+\delta Q\,,
\end{split}
\end{equation}
and expand
\begin{equation}
{\cal Q} =\bar{\cal Q} + \delta_1{\cal Q} + \delta_2{\cal Q}\,,
\end{equation}
where
\begin{align}
\delta_1{\cal Q} &= 2\left(\bar P-\alpha\bar Q\right)\delta P + 2\left(\beta\bar Q-\alpha\bar P\right)\delta Q\,,\\
\delta_2{\cal Q} &= (\delta P)^2 - 2\alpha\,\delta P\,\delta Q + \beta(\delta Q)^2\,.
\label{quadratic flux variation}
\end{align}
Combining the reduced flux action with the constant matter-vacuum contribution and using the background relation
\begin{equation}
\bar{\cal Q} = 2\kappa^2 \big(\alpha^2-\beta\big) V_{\rm vac}\,,
\label{background flux cancellation perturbations}
\end{equation}
the relevant part of the quadratic reduced action that is independent of the lapse fluctuation is, up to boundary terms,
\begin{equation}
\begin{split}
\delta_2S^{\rm flux+vac}_{\rm local} = \frac{1}{2\kappa^2} \int\dd y \int_{\Sigma_y}\dd^4x\,
\frac{1}{\alpha^2-\beta} \left[ \frac12 h\,\delta_1{\cal Q} + \delta_2{\cal Q} \right]\,.
\end{split}
\label{quadratic reduced flux action}
\end{equation}
The corresponding term proportional to the lapse fluctuation is included in~$\varphi\,{\cal C}^{(1)}$ in Eq.~\eqref{quadratic action decomposition}, rather than being counted separately in Eq.~\eqref{quadratic reduced flux action}.

The remaining matter contribution is denoted by~$\delta_2\Gamma_{\rm dyn}$. At the matter vacuum, the constant vacuum-energy term has already been included in deriving Eq.~\eqref{quadratic reduced flux action}. Intrinsic local curvature terms induced by matter loops are understood to have been absorbed into the renormalized gravitational couplings, while any remaining non-local form factors are neglected in the two-derivative stability analysis pursued here.

The linearised lapse constraint is
\begin{equation}
{\cal C}^{(1)}(y) = \int_{\Sigma_y}\dd^4x\, \Big[ \delta_1R^{(4)} - 2\big({\star_4\bar C_4}\big)\delta P - 2\big({\star_4\bar F_4}\big)\delta Q \Big]\,.
\label{linearised lapse constraint}
\end{equation}
To obtain this expression, we have used
\begin{equation}
\frac{\delta_1{\cal Q}}{2(\alpha^2-\beta)} = -\big({\star_4\bar C_4}\big)\delta P - \big({\star_4\bar F_4}\big)\delta Q\,,
\end{equation}
which follows from Eq.~\eqref{star F4 C4}. Because~${\cal C}^{(1)}(y)$ is itself an integral over four-dimensional spacetime, the projectable lapse imposes only a global constraint. In particular, it does not remove the local scalar polarisation associated with a non-Fierz–Pauli mass term for a non-zero Kaluza–Klein graviton.

Although a non-vanishing background flux therefore induces algebraic mixing between flux fluctuations and spacetime-averaged metric perturbations, this mixing is confined to the global sector. The variables~$\delta P$ and~$\delta Q$ should not be interpreted as ordinary four-dimensional scalar fields:~$\delta P$ is independent of~$y$, whereas~$\delta Q$ is independent of~$x^\mu$. Hence~$\delta P$ belongs entirely to the Kaluza–Klein zero-mode sector, while each Fourier coefficient of~$\delta Q$ is a spacetime-independent amplitude that couples only to the four-dimensional average of the metric trace.

We now specialise to the family of background solutions satisfying
\begin{equation}
\star_4\bar C_4=0\,.
\label{Cbar vanishing branch}
\end{equation}
Using Eq.~\eqref{star F4 C4}, this condition implies
\begin{equation}
\bar P=\alpha({\star_4\bar{F}_4})=\alpha\bar Q\,.
\end{equation}
The linearised modified Neumann condition~\eqref{deltaF} then gives
\begin{equation}
\delta Q(y)= \alpha\big({\star_4\bar C_4}\big)\varphi(y) = 0\,.
\label{deltaQ zero stability}
\end{equation}
Since~$Q$ is independent of~$x^\mu$, fixing~$\delta Q$ on the boundary of a slice~$\partial\Sigma_y$ fixes it over the entire slice. It follows that
\begin{equation}
\begin{split}
\delta_1{\cal Q}&=0\,,\\
\delta_2{\cal Q}&=(\delta P)^2\,.
\label{flux variations stability}
\end{split}
\end{equation}
Substituting Eq.~\eqref{flux variations stability} into Eq.~\eqref{quadratic reduced flux action} gives
\begin{equation}
\delta_2S_{\rm local}^{\rm flux+vac}  
= \frac{L}{2\kappa^2(\alpha^2-\beta)} \int\dd^4x\,(\delta P)^2\,,
\label{local flux action branch}
\end{equation}
where we have used the fact that~$\delta P(x)$ is independent of the compact coordinate. The remaining flux fluctuation therefore
belongs entirely to the Kaluza–Klein zero-mode sector and does not mix with any non-zero Fourier mode of the metric.\footnote{The branch~\eqref{Cbar vanishing branch} is a convenient family of sequestered backgrounds for the local stability analysis, although it is not the most general one. In particular, it satisfies
\begin{equation}
\bar{\cal Q} = \big(\beta-\alpha^2\big)\bar Q^2\,,
\end{equation}
and therefore corresponds, through Eq.~\eqref{background flux cancellation perturbations}, to~$V_{\rm vac}=-\frac{\bar Q^2}{2\kappa^2} \leq 0$ in the conventions adopted here. This restriction does not affect the local conclusion that higher-form perturbations do not alter the non-zero Kaluza–Klein graviton mass operator.}
In the absence of any additional zero-mode source or boundary condition fixing~$\delta P$, its equation of motion from~\eqref{local flux action branch} would further set~$\delta P=0$. This stronger statement is not needed for the non-zero-KK stability analysis.

Meanwhile, the linearised lapse constraint~\eqref{linearised lapse constraint} also simplifies to
\begin{equation}
{\cal C}^{(1)}(y) = \int_{\Sigma_y}\dd^4x\,\delta_1R^{(4)} \,.
\label{linear lapse branch}
\end{equation}
Since~$\delta_1R^{(4)} = \partial_\mu\partial_\nu h^{\mu\nu} - \Box h$ is a four-dimensional total derivative, the constraint~\eqref{linear lapse branch} is automatically satisfied for perturbations obeying the adopted boundary conditions. In particular, the projectable lapse imposes no local constraint capable of removing the scalar polarisation of a non-Fierz–Pauli massive graviton.

We have therefore established that, on the background branch~\eqref{Cbar vanishing branch}, the higher-form and projectable-lapse
sectors affect only the Kaluza–Klein zero mode and spacetime-averaged data. The local quadratic operator governing the non-zero
Kaluza–Klein gravitons is determined entirely by the gravitational action~\eqref{quadratic gravitational action}. We now turn to this local spectrum.\footnote{Although boundary terms and boundary conditions remain essential for the vacuum-energy cancellation mechanism, the analysis below concerns the local quadratic operator. Boundary conditions may restrict the global spectrum of admissible modes, but they do not change the local Fierz–Pauli tuning required to eliminate the additional scalar polarization.}

The fluctuation in the extrinsic curvature is
\begin{equation}
\delta_1 H_{\mu\nu} = \frac{1}{2} \big(\partial_y h_{\mu\nu} - 2\partial_{(\mu} n_{\nu)} \big)\,.
\label{linear extrinsic curvature}
\end{equation}
Since the background is flat and the compact direction is periodic, we can perform a Fourier decomposition of the metric and shift perturbations,
\begin{equation}
h_{\mu\nu}(x,y) = \sum_{n=-\infty}^{\infty} e^{2\pi {\rm i} n y/L}h_{\mu\nu}^{(n)}(x)\,,
    \qquad
    n_\mu(x,y) = \sum_{n=-\infty}^{\infty} e^{2\pi {\rm i} n y/L}n^{(n)}_\mu(x)\,.
\label{mode expansion}
\end{equation}
Reality conditions require that~$h^{(-n)}_{\mu\nu}(x) = h^{(n)*}_{\mu\nu}(x)$ and similarly for~$n^{(n)}_\mu$. 
For~$n\neq 0$, define
\begin{equation}
\xi_\mu^{(n)} \equiv \frac{1}{{\rm i}k_n}n_\mu^{(n)}\,,\qquad k_n\equiv\frac{2\pi n}{L}\,.
\label{Stuckelberg field definition}
\end{equation}
The corresponding gauge-invariant metric perturbation is
\begin{equation}
\hat{h}_{\mu\nu}^{(n)} \equiv h_{\mu\nu}^{(n)} - 2\partial_{(\mu}\xi_{\nu)}^{(n)}\,.
\label{gauge invariant KK metric}
\end{equation}
In terms of this combination,
\begin{equation}
\delta_1H_{\mu\nu}^{(n)} = \frac{{\rm i}k_n}{2} \hat{h}_{\mu\nu}^{(n)} = \frac{{\rm i}\pi n}{L} \hat{h}_{\mu\nu}^{(n)}\,.
\label{extrinsic curvature KK}
\end{equation}
Thus the non-zero shift modes provide the St\"uckelberg vector fields of the massive Kaluza–Klein gravitons, while~$\hat h_{\mu\nu}^{(n)}$ is the corresponding gauge-invariant spin-two field.

Substituting the mode expansions~\eqref{mode expansion} into the quadratic action~$\delta_2 S_{\text{grav}}$ and integrating over the fifth dimension yields
\begin{equation}
\begin{split}
\delta_2 S_{\rm grav} & = \frac{L}{2\kappa^2}  \int \dd^4x \left[ {\cal L}_{\rm EH}^{(2)}\big[h^{(0)}\big] - \lambda \partial_{(\mu}n_{\nu)}^{(0)}\partial^{(\mu}n^{(0)\nu)} + \mu \left( \partial_\mu n^{(0)\mu}\right)^2  \right]  \\
   &\quad
    + \sum_{n=1}^{\infty} \frac{L}{\kappa^2} \int \dd^4x \left[ {\cal L}_{\rm EH}^{(2)}\big[\hat{h}^{(n)}\big]
- \left(\frac{\pi n}{L}\right)^2 \left(\lambda\,\hat h_{\mu\nu}^{(n)}\hat h^{(n)*\mu\nu} - \mu\,\hat h^{(n)}\hat h^{(n)*} \right) \right]\\
    & \quad +\text{boundary terms}\,,
\end{split}
\label{quadratic graviton KK action}
\end{equation}
where~${\cal L}_{\rm EH}^{(2)}$ is the quadratic Einstein–Hilbert Lagrangian. In writing the non-zero-mode kinetic term as
${\cal L}_{\rm EH}^{(2)}[\hat{h}^{(n)}]$, we have used the gauge invariance of the quadratic Einstein–Hilbert action, up to a boundary term. The non-zero shift modes enter only through the gauge-invariant combination~$\hat{h}_{\mu\nu}^{(n)}$, as they should in their role as St\"uckelberg vector fields.

It is useful to consider separately the zero and non-zero Kaluza–Klein sectors. For the zero mode, the shift contribution in the first line of~\eqref{quadratic graviton KK action} may be rewritten as
\begin{equation}
-\lambda\,\partial_{(\mu}n_{\nu)}^{(0)} \partial^{(\mu}n^{(0)\nu)} +\mu\big(\partial_\mu n^{(0)\mu}\big)^2 =  -\frac{\lambda}{4} f_{\mu\nu}^{(0)}f^{(0)\mu\nu} + (\mu-\lambda) \big(\partial_\mu n^{(0)\mu}\big)^2\,,
\label{zero mode shift action}
\end{equation}
where~$f_{\mu\nu}^{(0)} \equiv 2\partial_{[\mu}n_{\nu]}^{(0)}$. For 
\begin{equation}
\lambda=\mu\,,
\label{lambda mu cond}
\end{equation}
the second term in Eq.~\eqref{zero mode shift action} vanishes, and the zero-mode shift has the familiar Maxwell form,~$-\frac{\lambda}{4} f_{\mu\nu}^{(0)}f^{(0)\mu\nu}$. This sector has a healthy kinetic sign provided~$\lambda>0$. Away from~$\lambda=\mu$, the longitudinal component of the zero-mode shift,~$n_\mu^{(0)} =\partial_\mu\phi$, acquires a four-derivative kinetic term,~$(\mu-\lambda) \big(\partial_\mu n^{(0)\mu}\big)^2 = (\mu-\lambda) (\Box\phi)^2$, which gives rise to a scalar ghost mode. 

The same condition~\eqref{lambda mu cond} follows independently from the non-zero Kaluza–Klein graviton modes. For every~$n\neq0$, the second line of Eq.~\eqref{quadratic graviton KK action} contains the four-dimensional spin-two mass term
\begin{equation}
{\cal L}_{\rm mass}^{(n)} = -\left(\frac{\pi n}{L}\right)^2 \left( \lambda\, \hat{h}_{\mu\nu}^{(n)} \hat{h}^{(n)*\mu\nu}
- \mu\, \hat{h}^{(n)}\hat{h}^{(n)*} \right)\,.
\label{KK graviton mass term}
\end{equation}
A generic Lorentz-invariant mass term for a four-dimensional spin-two field propagates a scalar mode with a wrong-sign kinetic term (in addition to the five massive spin-two polarisations). This additional degree of freedom is absent only for the Fierz–Pauli combination~\cite{Fierz:1939ix}, which in the present notation again requires~$\lambda=\mu$. At this point the mass term becomes
\begin{equation}
{\cal L}_{\rm mass}^{(n)} = -\lambda \left(\frac{\pi n}{L}\right)^2 \left( \hat{h}_{\mu\nu}^{(n)} \hat{h}^{(n)*\mu\nu}
- \hat{h}^{(n)}\hat{h}^{(n)*} \right)\,,
\end{equation}
and the spin-two Kaluza–Klein masses are
\begin{equation}
m_n^2 = 2\lambda \left(\frac{\pi n}{L}\right)^2 
\label{KK graviton masses}
\end{equation}
with $n\geq1$ and the normalization adopted in Eq.~\eqref{quadratic graviton KK action}. Stability of the massive spin-two spectrum therefore additionally requires~$\lambda>0$.\footnote{Away from~$\lambda = \mu$, it is sufficient for any ghost degree of freedom to lie above the cut-off of the effective theory. For the~$n$\textsuperscript{th} Kaluza--Klein mode, the ghost mass squared scales as~\cite{VanNieuwenhuizen:1973fi,Hinterbichler:2011tt}
\begin{equation}
    \left( m^{(n)}_{\text{ghost}} \right)^2
    =
    \frac{\lambda(4\mu - \lambda)}{\lambda - \mu}
    \left( \frac{\pi n}{L} \right)^2
\end{equation}
for $n\geq 1$, which diverges in the Fierz--Pauli limit. This analysis and result is analogous to the one performed in projectable Ho\v{r}ava gravity \cite{Blas:2010hb}. There, the preferred normal direction is timelike and the scalar mode manifests itself through a gradient instability. The physical interpretation here is different in that we find a non-Fierz--Pauli scalar ghost of a four-dimensional massive Kaluza--Klein graviton.}

The condition~$\lambda=\mu$ should not by itself be confused with full five-dimensional Lorentz invariance. It guarantees the Fierz–Pauli tensor structure of the non-zero Kaluza–Klein masses and the Maxwell form of the zero-mode shift at quadratic order. The five-dimensional Einstein–Hilbert point of the gravitational bulk sector is the particular choice
\begin{equation}
\lambda=\mu=1\,.
\end{equation}
For other positive values satisfying~$\lambda=\mu$, the linearised spectrum has the same healthy tensor structure, although the full non-linear theory does not possess the complete five-dimensional diffeomorphism symmetry of Einstein gravity.

In summary, the projectable lapse and higher-form fields constrain only spacetime-averaged flux and metric fluctuations; they do not remove the local scalar polarisation generated by a non-Fierz–Pauli mass term. For each non-zero Kaluza–Klein mode, absence of this scalar ghost requires the Fierz–Pauli relation~$\lambda=\mu$. Independently, the same relation reduces the zero-mode shift sector to a Maxwell action, whereas for~$\lambda\neq\mu$ its longitudinal component obeys a fourth-order equation and carries an additional ghost. The natural linearly stable branch of the complete periodic theory is therefore
\begin{equation}
\lambda=\mu>0\,,
\end{equation}
which includes the five-dimensional Einstein–Hilbert point~$\lambda=\mu=1$. The higher-form sector remains essential for the global cancellation of the vacuum energy, but it neither changes nor relaxes this local stability condition.\footnote{Our analysis is restricted to the quadratic action. At this order, the condition~$\lambda=\mu$ ensures that the non-zero Kaluza–Klein gravitons have the Fierz–Pauli tensor structure, while simultaneously reducing the zero-mode shift sector to Maxwell form. Whether these properties persist non-linearly is a separate question. In particular, one should verify that non-linear interactions do not reintroduce the additional scalar degree of freedom, analogous to the Boulware–Deser mode of non-linear massive gravity. We leave such an analysis for future work.}

\section{A covariant formulation from a spacelike khoron} \label{sec:stuck}

We have shown that the inclusion of bulk dynamics for both gravity and fluxes circumvents the gravitational effect of vacuum energy through the interplay of four-form fluxes, the lapse constraint, and periodic boundary conditions. Owing to the projectability condition, the lapse constraint is local along the compact extra dimension but global over each four-dimensional spacetime slice. Although this formulation is consistent with four-dimensional Lorentz invariance, its foliation-based language is somewhat unfamiliar. In this section, we recast the mechanism in a manifestly five-dimensional diffeomorphism-covariant form. This is achieved through a St\"uckelberg completion of the preferred foliation: rather than fixing the foliation at the outset and retaining only foliation-preserving diffeomorphisms, we describe it covariantly by a scalar field.

\subsection{Covariant action and projectability}

Our aim is to formulate the mechanism in a manifestly covariant language and thereby gain a deeper understanding of the vacuum energy cancellation mechanism. 
Motivated by the ghost-free condition found in the previous section, we specialize the gravitational and higher-form bulk sectors to their five-dimensional generally covariant form,
\begin{equation}
   S_{\rm grav + flux} = \frac{1}{2\kappa^2} \int\dd^5X\,\sqrt{-\gamma}\,R^{(5)}(\gamma) - \frac{1}{2\kappa^2} \int\mathcal F_5\wedge\star_5\mathcal F_5  + \text{boundary terms}\,, 
\label{action5D grav}
\end{equation}
where~$R^{(5)}(\gamma)$ is the Ricci scalar of the 5d metric~$\gamma_{ab}$. The four-form potential~$\mathcal{B}_4$ gives rise to the 5-form field strength~$\mathcal{F}_5=\dd_5 \mathcal{B}_4$. 

The preferred foliation is defined via a spacelike scalar, befittingly referred to as a \emph{khoron}\footnote{{\it Khora} is the ancient Greek term for space, and so the khoron is the spacelike analogue of the well-known khronon field familiar from Lorentz-violating theories such as Hořava–Lifshitz gravity.} and denoted~$ Y(X)$, whose level sets are the four-dimensional leaves of the foliation. Since the fifth dimension is compact,~$ Y$ should be regarded as a periodic scalar. We work in the unit-winding sector and assume throughout that~$\partial_a Y$ is everywhere non-vanishing, so that its level sets define a smooth foliation and unitary gauge~$ Y=y$ may be imposed globally. The khoron determines the unit spacelike normal
\begin{equation}
n_a(X)=\frac{1}{{\cal X}}\partial_a Y \,,
\label{khoron normal}
\end{equation}
where
\begin{equation}
{\cal X}\equiv \sqrt{\gamma^{ab}\partial_a Y\partial_b Y}\,,
\end{equation}
and~$\gamma_{ab}$ is the bulk five-dimensional metric. The foliation is invariant under orientation-preserving reparameterizations,~$ Y \rightarrow \widetilde{ Y}( Y)$ with~$\widetilde{ Y}'( Y)> 0$,
under which~$n_a$ is invariant. The induced metric on each leaf is
\begin{equation}
q_{ab}=\gamma_{ab}-n_an_b\,.
\end{equation}
The extrinsic curvature of the leaves and its trace are
\begin{equation}
H_{ab} = \tensor{q}{_a^{c}}\tensor{q}{_b^{d}} D_cn_d\,, \qquad H = q^{ab}H_{ab} = D_an^a\,,
\label{Hn}
\end{equation}
where~$D_a$ is the covariant derivative associated with~$\gamma_{ab}$.

The projectable limit is equivalent to requiring that the khoron congruence be geodesic,
\begin{equation}
n^bD_bn^a=0\,.
\label{geodesic for n_a}
\end{equation}
Indeed, using~\eqref{khoron normal}, its acceleration can be written as
\begin{equation}
n^bD_bn^a=q^{ab}D_b\ln{\cal X}\,.
\label{geodesic for n_a 2}
\end{equation}
Thus the geodesic condition is equivalent to requiring that~${\cal X}$ be constant along each leaf. To make contact with the original formulation, consider the ADM decomposition~\eqref{metric}, where the lapse and shift initially depend on all coordinates~$X^a=(x^\mu,y)$. In unitary gauge,~$ Y=y$, one has~${\cal X}=1/N$. Equation~\eqref{geodesic for n_a 2} then implies~$\partial_\mu N=0$, or equivalently~$N=N(y)$, reproducing the projectable limit.

Since the projectable condition is equivalent to requiring that~${\cal X}$ be constant on each leaf, we now impose this condition covariantly. Let~$\sigma$ denote a coordinate on the one-dimensional space of leaves, with the leaf~$\Sigma_\sigma$ defined by~$ Y(X)=\sigma$, and introduce an independent positive auxiliary einbein~$\eta(\sigma)$ on leaf space. Under a reparameterization of the leaf coordinate,
\begin{equation}
\widetilde\eta(\widetilde\sigma)\,\dd\widetilde\sigma = \eta(\sigma)\,\dd\sigma\,.
\end{equation}
Its pullback to spacetime is denoted by~$\eta( Y(X))$. The condition we impose is
\begin{equation}
{\cal X}(X)=\frac{1}{\eta\big( Y(X)\big)}\,.
\label{X rho}
\end{equation}
The introduction of~$\eta$ is required for this relation to be covariant under reparameterizations of the khoron. Indeed, under an orientation-preserving transformation~$ Y\rightarrow\widetilde{ Y}( Y)$,
\begin{equation}
\widetilde{\cal X} =\widetilde{  Y}'( Y)\,{\cal X}\,, \qquad \widetilde{\eta}(\widetilde{  Y}) = \frac{\eta( Y)}{\widetilde{ Y}'( Y)}\,,
\end{equation}
so that~$\eta( Y){\cal X}$ is invariant. Since~$\eta( Y)$ is the pullback of a function on the one-dimensional space of leaves, it is constant on every leaf. Eq.~\eqref{X rho} therefore implies~$q^{ab}D_b{\cal X}=0$, and thus, by Eq.~\eqref{geodesic for n_a 2}, also implies the geodesic condition~\eqref{geodesic for n_a}. We enforce~\eqref{X rho} with a scalar Lagrange multiplier~$\Xi(X)$ through the action
\begin{equation}
S_\Xi = \int \dd^5X\sqrt{-\gamma}\, \Xi(X) \Big[ \eta\big( Y(X)\big){\cal X}(X)-1 \Big]\,.
\label{constraint action}
\end{equation}
Variation with respect to~$\Xi$ reproduces the constraint~\eqref{X rho}.

The covariantization of the matter action requires some care because of our working assumption that it is ultra-local in the extra dimension. The corresponding expression should reduce in unitary gauge to
\begin{equation}
S_{\rm m} = \int \dd y\, N(y)\, \Gamma_{\rm m}[g_y,\Psi_y] \qquad \text{(unitary gauge)} \,.
\label{Sm unitary}
\end{equation}
To express this covariantly, it is convenient to regard the theory as assigning a four-dimensional effective-action functional to each leaf of the foliation. We take the matter fields~$\Psi(X)$ to be defined throughout the five-dimensional spacetime, and denote their pullbacks to the leaf~$\Sigma_\sigma$, together with the induced metric, by~$(q_\sigma,\Psi_\sigma)$. Accordingly, the matter dynamics is described by a single four-dimensional effective-action functional~$\Gamma_{\rm m}[q,\Psi]$, evaluated on the induced fields of each leaf, $\Gamma_{\rm m}[q_\sigma,\Psi_\sigma]$. By assumption, this functional may be arbitrarily non-local within a given leaf, but it contains no couplings between fields on distinct leaves. In this sense, the matter theory is ultra-local in the direction normal to the foliation. A general off-shell configuration may nevertheless vary from one leaf to another.

The same einbein introduced above provides the natural measure over the foliation. We therefore define
\begin{equation}
S_{\rm m} = \int\dd\sigma\, \eta(\sigma)\, \Gamma_{\rm m}[q_\sigma,\Psi_\sigma]\,.
\label{leaf matter action}
\end{equation}
The constraint~\eqref{X rho} identifies this measure with the proper-distance measure induced by the five-dimensional geometry. In unitary gauge, where~${\cal X}=1/N$, it gives~$\eta(y)=N(y)$, and~\eqref{leaf matter action} reduces immediately to~\eqref{Sm unitary}.\footnote{It may be tempting to define the matter action directly as
\begin{equation}
S_{\rm m} = \int \frac{\dd Y}{{\cal X}} \Gamma_{\rm m}[q_ Y,\Psi_ Y] \,.
\end{equation}
The issue is that, unless the geodesic condition~\eqref{geodesic for n_a} holds,~${\cal X}$ generally varies across a given leaf. Its inverse is then a point-dependent spacetime scalar rather than a well-defined function on the space of leaves. Since~$\Gamma_{\rm m}[q_ Y,\Psi_ Y]$ is a single functional associated with the entire leaf (and may itself be non-local within that leaf), there is no unambiguous prescription for multiplying it by the point-dependent quantity~$1/{\cal X}$.

When the matter theory admits a local representation in terms of fields intrinsic to the leaves and their tangential derivatives, Eq.~\eqref{leaf matter action} can instead be written as
\begin{equation}
S_{\rm m} = \int \dd\sigma\, \eta(\sigma) \int_{\Sigma_\sigma} \dd^4x \sqrt{-q} \,{\cal L}_{\rm m}\big(q,\Psi,\nabla\Psi,\ldots\big) \,.
\end{equation}
By the coarea formula (see Eq.~\eqref{coarea} below), this becomes
\begin{equation}
S_{\rm m} = \int \dd^5X\sqrt{-\gamma} \,\eta( Y) {\cal X} \, {\cal L}_{\rm m}\,,
\end{equation}
which on the constraint surface,~$\eta {\cal X} = 1$, reduces to~$S_{\rm m} = \int \dd^5X\sqrt{-\gamma} \, {\cal L}_{\rm m}$. Thus a local leaf theory admits an ordinary five-dimensional integral representation on the constraint surface, while Eq.~\eqref{leaf matter action} provides the appropriate off-shell completion needed for the variational principle.}

The full action is the sum of Eqs.~\eqref{action5D grav},~\eqref{constraint action} and~\eqref{leaf matter action}:
\begin{equation}
\begin{split}
   S & = \frac{1}{2\kappa^2} \int\dd^5X\,\sqrt{-\gamma}\,R^{(5)}(\gamma) - \frac{1}{2\kappa^2} \int\mathcal F_5\wedge\star_5\mathcal F_5    \\
  & + \int\dd\sigma\, \eta(\sigma)\, \Gamma_{\rm m}[q_\sigma,\Psi_\sigma] + \int \dd^5X \sqrt{-\gamma} \,\Xi(X)\Big[\eta\big( Y(X)\big){\cal X}(X) -1 \Big] + \text{boundary terms}\,.
\end{split}
\label{action5D}
\end{equation}
Variation of the four-form potential forces a constant 5-form flux, which we write as
\begin{equation}
\mathcal{F}_{abcde}=f\epsilon_{abcde}\,,
\end{equation}
where~$f$ is an integration constant, and~$\epsilon_{abcde}$ is the 5d Levi-Civita tensor. As with the constraint~\eqref{X rho}, the flux solution is imposed only after varying the action with respect to the metric. 

Although the einbein~$\eta$ is held fixed when varying~$ Y$ and~$\gamma_{ab}$, it is itself varied independently and acts as a Lagrange-multiplier-like variable enforcing a leaf-wise global constraint.
To vary the action with respect to~$\eta(\sigma)$, it is useful to rewrite the constraint action~\eqref{constraint action} using the coarea formula,\footnote{To prove Eq.~\eqref{coarea}, choose coordinates~$(x^\mu,\sigma)$ adapted to the foliation, with~$\sigma= Y(X)$. In these coordinates the ADM lapse in the direction normal to the leaves is~$1/{\cal X}$, and therefore~$\sqrt{-\gamma} = \sqrt{-q}/{\cal X}$. It follows immediately that
\begin{equation}
\dd^5X\,\sqrt{-\gamma} = \dd\sigma\,\dd^4x\, \frac{\sqrt{-q}}{{\cal X}}\,,
\end{equation}
which gives~\eqref{coarea}.}
\begin{equation}
\int \dd^5X\,\sqrt{-\gamma}\,F(X) = \int \dd\sigma \int_{\Sigma_\sigma}\dd^4x\, \frac{\sqrt{-q}}{{\cal X}}\,F(X)\,,
\label{coarea}
\end{equation}
such that
\begin{equation}
\int \dd^5X \sqrt{-\gamma}\, \eta\big( Y(X)\big) \,\Xi(X) {\cal X}(X)   =  \int \dd\sigma\,\eta(\sigma) \int_{\Sigma_\sigma}\dd^4x\sqrt{-q}\,\Xi(X) \,.
\end{equation}
The einbein equation is consequently 
\begin{equation}
\big\langle\Xi\big\rangle_\sigma = -\frac{\Gamma_{\rm m}[q_\sigma,\Psi_\sigma]} {{\cal V}_4(\sigma)}\,,
\label{rho equation avg}
\end{equation} 
where the regulated spacetime average on a leaf and four-volume are defined analogously to Eq.~\eqref{O avg}, 
\begin{equation}
  \langle \mathcal{O}\rangle_\sigma 
  = \frac{1}{\mathcal{V}_4(\sigma)} \int_{\Sigma_\sigma} \dd^4x \sqrt{-q}\, \mathcal{O}\,, \qquad \mathcal{V}_4(\sigma)=\int_{\Sigma_\sigma} \dd^4x \sqrt{-q}\,.
\label{O avgsig}
\end{equation}
Thus the einbein equation gives a global constraint on the scalar Lagrange multiplier.

\subsection{Khoron equation of motion}

To obtain the khoron equation of motion, we vary the action~\eqref{action5D} with respect to~$ Y$, holding the bulk metric and $p$-form fields fixed, along with the Lagrange multiplier $\Xi$ and the leaf-space einbein~$\eta(\sigma)$. On each leaf, the induced metric, $q^\sigma_{ab}$, and the pullback of the matter fields, $\Psi_\sigma$, pick up implicit variations due to the dependence on the khoron. It follows that only the constraint action and the matter action contribute directly to the khoron equations of motion.

Varying~$ Y$ at fixed value of~$\sigma$ changes which spacetime points belong to that hypersurface. In other words, we let $ Y \to \widetilde{ Y}= Y+\delta  Y$ and correspondingly $X^a \to \widetilde X^a=X^a+\delta X^a$ such that $ Y(X)=\widetilde{ Y} (\widetilde X)=\sigma$. To first order, this gives,
\begin{equation}
\delta  Y+\delta X^a\partial_a Y=0 \,.
\end{equation}
Since $\partial_a  Y$ points along the normal, this variation only cares about a normal displacement, 
\begin{equation}
\delta X^a=\zeta n^a\,.
\label{delta Xa}
\end{equation}
Using~$n^a\partial_a  Y = {\cal X}$, we can relate the displacement in $ Y$ to the corresponding coordinate displacement along the normal, 
\begin{equation}
\delta  Y=-\zeta \cal X \,.
\label{zeta def}
\end{equation}

\vspace{0.2cm}
\noindent {\bf Variation of the constraint action:} We begin by varying the constraint action~$S_\Xi$ with respect to the khoron. Since~$\eta$ is held fixed as a function on leaf space,
\begin{equation}
\delta\big(\eta( Y) {\cal X} \big) = \eta’( Y){\cal X} \;\delta  Y + \eta( Y)n^aD_a\delta  Y\,,
\end{equation}
where we have used~$\delta{\cal X} = n^aD_a\delta  Y$. Thus
\begin{equation}
\delta S_\Xi = \int\dd^5X\,\sqrt{-\gamma}\, \left[ \Xi(X) \eta’ ( Y){\cal X}  - D_a\Big(\Xi(X) \eta( Y) n^a\Big) \right]\delta  Y \,,
\end{equation}
where we have integrated the second term by parts and dropped boundary terms. Since~$\eta = \eta( Y)$, we have
\begin{equation}
D_a\Big(\Xi(X) \eta( Y) n^a\Big) = \eta( Y) D_a\big(\Xi(X) n^a\big) + \Xi(X)  \eta’ ( Y){\cal X} \,.
\end{equation}
The terms proportional to~$\eta'$ cancel out of~$\delta S_\Xi$, leaving us with
\begin{equation}
\begin{split}
\delta S_\Xi &= -\int\dd^5X\,\sqrt{-\gamma}\, \eta( Y) D_a\big(\Xi(X) n^a\big) \delta  Y\\
    &= -\int\dd^5X\,\sqrt{-\gamma}\, \eta( Y) \Big[D_n\Xi + H \Xi \Big] \delta  Y \,.
\label{dSXi dY}
\end{split}
\end{equation} 
Recall that~$D_n = n^a D_a$ and~$H = D_a n^a$ is the trace of the extrinsic curvature. 

\vspace{0.2cm}
\noindent {\bf Variation of the matter action:} We now consider the matter action~$S_{\rm m}$. The matter functional~$\Gamma_{\rm m}[q_\sigma,\Psi_\sigma]$ is associated with the embedded hypersurface~$\Sigma_\sigma$. Since $\sigma$ and $\eta$ are held fixed, variation of the matter action only comes of the khoron dependence in the induced metric and the matter fields. Under the embedding deformation~\eqref{delta Xa}, an ambient field pulled back to the leaf varies by its Lie derivative along~$\zeta n^a$,
\begin{equation}
\delta q^\sigma_{ab} = 2\zeta H_{ab}\,,\qquad \delta \Psi_\sigma = \zeta {\cal L}_n\Psi \,.
\end{equation}
Here~$H_{ab}$ is the extrinsic curvature defined in Eq.~\eqref{Hn}, and~${\cal L}_n\Psi$ denotes the appropriate Lie derivative of the pulled-back matter fields (reducing to~$D_n\Psi$ for a scalar). The appearance of~${\cal L}_n\Psi$ does not contradict ultra-locality: the latter means that the action contains no normal-derivative couplings, not that a general off-shell field configuration must be constant from one leaf to the next.

The variation of the matter action now yields

\begin{align}
\delta S_{\rm m} &=  \int \dd\sigma\,\eta(\sigma) \int_{\Sigma_\sigma}\dd^4x\,\sqrt{-q} \left[\frac12 \tau^{(\rm m)ab}\delta q^\sigma_{ab} + \frac{1}{\sqrt{-q}} \frac{\delta\Gamma_{\rm m}}{\delta\Psi} \delta\Psi_\sigma  \right] \\
&=  \int \dd\sigma\,\eta(\sigma) \int_{\Sigma_\sigma}\dd^4x\,\sqrt{-q} \left[\tau^{(\rm m)ab} H_{ab} + \frac{1}{\sqrt{-q}} \frac{\delta\Gamma_{\rm m}}{\delta\Psi} {\cal L}_n\Psi  \right]\zeta \\
&= - \int \dd^5 X\sqrt{-\gamma}\, \eta( Y) \left[\tau^{(\rm m)ab} H_{ab} + \frac{1}{\sqrt{-q}} \frac{\delta\Gamma_{\rm m}}{\delta\Psi} {\cal L}_n\Psi  \right]\delta  Y \,, \label{dSm dYexp}
\end{align}
where we have defined the intrinsic matter stress tensor on each leaf by
\begin{equation}
\tau^{(\rm m)}_{ab} \equiv -\frac{2}{\sqrt{-q}} \frac{\delta\Gamma_{\rm m}}{\delta q^{ab}}\,,
\qquad
n^a\tau^{(\rm m)}_{ab}=0\,.
\label{tau matter def}
\end{equation}
To arrive at the last line above we have made use of Eq.~\eqref{zeta def} along with the coarea law \eqref{coarea}. 

Combining Eqs.~\eqref{dSXi dY} and~\eqref{dSm dYexp}, and using the fact that~$\eta\neq 0$ in the regular foliation sector, the khoron equation of motion is
\begin{equation}
D_n\Xi+H\Xi=-\left[\tau^{(\rm m)ab} H_{ab} + \frac{1}{\sqrt{-q}} \frac{\delta\Gamma_{\rm m}}{\delta\Psi} {\cal L}_n\Psi  \right]\,.
\label{khoron eom for Xi}
\end{equation}
Of course, on the matter equations of motion, $\frac{\delta\Gamma_{\rm m}}{\delta\Psi}=0$, so this equation tells us that normal evolution of $\Xi$ is determined exclusively by the matter stress energy on each leaf. Its spacetime average on each leaf is fixed independently by the einbein equation~\eqref{rho equation avg}. 

\subsection{Einstein equations}

We next derive the five-dimensional Einstein equations and project them onto the leaves of the khoron foliation. In deriving the metric equations, both the projectability constraint~\eqref{X rho} and the five-form solution are imposed only after the corresponding variations have been performed. In particular, the leaf-space einbein~$\eta(\sigma)$ and the khoron~$ Y$ are both held fixed when varying~$\gamma_{ab}$.

Variation of the action with respect to the metric yields
\begin{equation}
G^{(5)}_{ab}(\gamma)=-\frac{1}{2} f^2 \gamma_{ab} + \kappa^2 T_{ab}^{(\Xi)} + \kappa^2 T^{(\rm m)}_{ab}\,.
\label{Einstein 5d} 
\end{equation}
As usual for a top form, the five-form flux gravitates as a five-dimensional cosmological constant. We derive the remaining stress-tensor contributions in turn.

\vspace{0.2cm}
\noindent {\bf Stress tensor of the constraint sector:} Consider first the constraint action~\eqref{constraint action}. Using
\begin{equation}
\begin{split}
\delta{\cal X} &= \frac{1}{2{\cal X}}\partial_a Y\partial_b Y\,\delta\gamma^{ab} \\
&=\frac{{\cal X}}{2}n_an_b\,\delta\gamma^{ab}\,,
\end{split}
\end{equation}
we obtain
\begin{equation}
\begin{split}
T^{(\Xi)}_{ab} &= -\frac{2}{\sqrt{-\gamma}}\frac{\delta S_\Xi}{\delta\gamma^{ab}}\\
&= \Xi\big(\eta{\cal X}-1\big)\gamma_{ab} -\Xi\eta{\cal X}\,n_an_b\,.
\label{TXi offshell}
\end{split}
\end{equation}
On the constraint surface,~$\eta{\cal X}=1$, this reduces to
\begin{equation}
T^{(\Xi)}_{ab} =-\Xi\, n_a n_b\,.
\label{TXi onshell}
\end{equation}
Thus the constraint sector stress tensor is purely normal. 

\vspace{0.2cm}
\noindent {\bf Matter stress tensor:} We now derive the variation of the matter action with respect to the metric, $\gamma_{ab}$. At fixed~$ Y$, using ${\delta n_a = \frac{1}{2} n_a n^c n^d\delta\gamma_{cd}}$, the variation of the induced metric is given by
\begin{equation}
\delta q_{ab} = \delta \gamma_{ab} - n_a n_b n^c n^d\delta\gamma_{cd}\,.
\label{del q del gamma}
\end{equation} 
We also recall our earlier definition of the intrinsic matter stress tensor on each leaf \eqref{tau matter def}. Variation of the matter action now gives
\begin{equation}
\delta S_{\rm m} =  \frac{1}{2}\int \dd\sigma\,\eta(\sigma) \int_{\Sigma_\sigma}\dd^4x\sqrt{-q}\, \tau^{(\rm m)ab}\delta \gamma_{ab} \,,
\end{equation}
where we have used Eq.~\eqref{del q del gamma} and~$n^a\tau^{(\rm m)}_{ab}=0$. Equivalently, using the coarea formula~\eqref{coarea},
\begin{equation}
\delta S_{\rm m} = \frac{1}{2}\int \dd^5X\sqrt{-\gamma}\, \eta\big( Y(X)\big) {\cal X}(X) \,\tau^{(\rm m)ab}\,\delta\gamma_{ab}\,.
\end{equation}
The five-dimensional stress tensor associated with~$S_{\rm m}$ is therefore
\begin{equation}
T^{(\rm m)}_{ab} =  \eta{\cal X}\tau^{(\rm m)}_{ab}\,.
\label{5d matter stress}
\end{equation}
The factor~$\eta{\cal X}$ is precisely the off-shell factor appearing when a local leaf action is rewritten using the coarea formula.
On the constraint surface~$\eta{\cal X}=1$, this becomes simply
\begin{equation}
T^{(\rm m)}_{ab}=\tau^{(\rm m)}_{ab}\,.
\label{matter tangent}
\end{equation}
In particular,~$n^aT^{(\rm m)}_{ab} = 0$, and the matter stress tensor is purely tangential. This does not mean that matter drops out of the normal gravitational constraint. Its normal gravitational response is carried indirectly by~$\Xi$, whose leaf average is tied to the complete matter effective action through Eq.~\eqref{rho equation avg}.

Substituting~\eqref{TXi onshell} and~\eqref{matter tangent}, the Einstein equations~\eqref{Einstein 5d} become
\begin{equation}
G^{(5)}_{ab} =-\frac{1}{2} f^2 \gamma_{ab} -\kappa^2 \Xi\, n_a n_b + \kappa^2 \tau^{(\rm m)}_{ab}\,.
\label{5d Einstein Xi}
\end{equation}
We now decompose these equations into components tangential and normal to the leaf using the normal vector~$n_a$ and the induced metric~$q_{ab}=\gamma_{ab}-n_an_b$, whose covariant derivative will be denoted by~$\nabla_a$. All four-dimensional tensors below are intrinsic to a leaf, with tangential indices implicitly projected using the induced metric.

For a geodesic spacelike normal, the Gauss–Codazzi relations imply
\begin{equation}
q^{ac} \tensor{q}{_b^d} G^{(5)}_{cd} = \tensor{G}{^{(4)a}_{b}} -\kappa^2\tensor{T}{^{(H)a}_{b}}\,,
\label{Gauss tangential}
\end{equation}
where
\begin{equation}
\tensor{T}{^{(H)a}_{b}} =  \frac{1}{\kappa^2} \Big[ H\tensor{H}{^a_b} - \frac{1}{2} \tensor{q}{^a_b} \big(H_{cd}H^{cd}+H^2\big) +{\cal L}_n \big(\tensor{H}{^a_b}-H\tensor{q}{^a_b} \big) \Big]\,.
\label{TH covariant}
\end{equation}
This is the covariant version of Eq.~\eqref{TH} with~$\lambda = \mu = 1$. Using Eq.~\eqref{5d Einstein Xi}, the tangential projection becomes
\begin{equation}
\tensor{G}{^{(4)a}_b} =  -\frac{1}{2}f^2 \tensor{q}{^a_b} +\kappa^2\tensor{T}{^{(H)a}_b} +\kappa^2 \tensor{\tau}{^{(\rm m)a}_b}\,.
\label{G4 Xi}
\end{equation}
The normal-normal projection is
\begin{equation}
n^an^b G^{(5)}_{ab} =  -\frac{1}{2} \left( R^{(4)}+H_{ab}H^{ab}-H^2 \right)\,.
\label{Gauss nn}
\end{equation}
Substituting Eq.~\eqref{5d Einstein Xi} and using the fact that~$\tau^{(\rm m)}_{ab}$ is purely tangential, we obtain
\begin{equation}
R^{(4)} =  f^2-H_{ab}H^{ab}+H^2+2\kappa^2\Xi\,.
\label{R4 Xi}
\end{equation}
Lastly, the mixed projection gives the Codazzi constraint
\begin{equation}
\tensor{q}{_a^c}n^dG^{(5)}_{cd} =  \nabla_b\tensor{H}{^b_a}-\nabla_aH\,.
\label{Gauss mixed}
\end{equation}
Since neither~$\tau^{(\rm m)}_{ab}$ nor~$T^{(\Xi)}_{ab}$ has a mixed projection, this yields
\begin{equation}
\nabla_b\tensor{H}{^b_a}-\nabla_aH = 0\,.
\label{momentum constraint Xi}
\end{equation}
Taking the four-dimensional trace of Eq.~\eqref{G4 Xi} and using Eq.~\eqref{TH covariant}, one obtains
\begin{equation}
R^{(4)} = 2f^2+ H^2 + 2H_{ab}H^{ab} + 3{\cal L}_nH  - \kappa^2\tau^{(\rm m)}\,,
\label{R4 from trace}
\end{equation}
where~$\tau^{(\rm m)} = q^{ab} \tau^{(\rm m)}_{ab}$. Eliminating~$R^{(4)}$ using Eq.~\eqref{R4 Xi} gives
\begin{equation}
f^2 = -3 \big(H_{ab}H^{ab} + {\cal L}_nH\big) +2\kappa^2\Xi +\kappa^2\tau^{(\rm m)} \,.
\label{f equation Xi}
\end{equation}
Eq.~\eqref{f equation Xi} holds pointwise, while the four-form equation independently requires~$f$ to be a single spacetime constant. Consequently its right-hand side must be constant on every solution. Averaging the equation over a four-dimensional leaf therefore loses no information about~$f$, but makes explicit the global constraint needed to eliminate it from the tangential Einstein equation. Using the einbein equation~\eqref{rho equation avg}, we obtain
\begin{equation}
f^2 = -3 \left\langle H_{ab}H^{ab} + {\cal L}_nH \right\rangle_\sigma + \kappa^2 \big\langle \tau^{(\rm m)}\big\rangle_\sigma - 2 \kappa^2 \frac{\Gamma_{\rm m}[q_\sigma,\Psi_\sigma]}{{\cal V}_4(\sigma)}\,.
\label{f average Xi}
\end{equation}
Substituting this into the tangential Einstein equations~\eqref{G4 Xi} gives
\begin{equation}
\begin{split}
\tensor{G}{^{(4)a}_{b}}
= \kappa^2 \tensor{T}{^{(H)a}_b} &+ \frac{3}{2} \tensor{q}{^a_b}  \left\langle H_{cd}H^{cd} + {\cal L}_nH \right\rangle_\sigma   \\
&+\kappa^2\left(\tensor{\tau}{^{({\rm m})a}_{b}} -\frac{1}{2} \tensor{q}{^a_b} \big\langle \tau^{(\rm m)}\big\rangle_\sigma\right) + \kappa^2 \frac{\Gamma_{\rm m}[q_\sigma,\Psi_\sigma]}{{\cal V}_4(\sigma)} \, \tensor{q}{^a_b} \,.
\end{split}
\label{Ein khoron effective}
\end{equation}
This recovers the corresponding version of the effective Einstein equation \eqref{Ein2}, written in covariant form. Here, however, we have $\lambda=\mu=1$ and there is no local fluctuation in the flux since $f$ is everywhere constant. Of course, this expression still contains Lie derivatives of the extrinsic curvature. 
 As in Sec.~\ref{sec:def}, we can eliminate these by averaging over the compact dimension (see for instance Eqs.~\eqref{eq:THmunu_average}--\eqref{Ein3}), with the $S^1$ average defined in \eqref{weightedav} now written using the einbein, 
\begin{equation} \label{weightedavsig}
  \langle \mathcal{O} \rangle_{S^1}
  = \frac{\int_{S^1} \dd \sigma \,\eta(\sigma) \,\mathcal{O}}
         {\int_{S^1} \dd \sigma \eta(\sigma) }\,.
\end{equation}

It is worth pausing to interpret the role of~$\Xi$ in this result. The multiplier does not appear as an additional local source in the tangential Einstein equation~\eqref{G4 Xi}; its stress tensor is entirely normal to the foliation. Nevertheless, its leaf average enters crucially through the normal gravitational constraint. The einbein equation~\eqref{rho equation avg} fixes this leaf-wise global component,~$\langle\Xi\rangle_\sigma = -\Gamma_{\rm m}[q_\sigma,\Psi_\sigma]/{\cal V}_4(\sigma)$, while the khoron equation~\eqref{khoron eom for Xi},~$D_n\Xi+H\Xi=-\tau^{ab}_{\rm m}H_{ab}$, determines its evolution in the normal direction. Thus~$\Xi$ plays two complementary roles: its leaf average implements the global constraint inherited from the projectable lapse, whereas its local profile ensures that this constraint is embedded consistently in a manifestly five-dimensional covariant theory.

This also clarifies the status of~$f$. Eq.~\eqref{f equation Xi} should not be interpreted as allowing~$f$ to be chosen independently at every spacetime point. The four-form equation first establishes that~$f$ is constant; the Einstein and einbein equations then determine the value of that constant as part of the global solution. In this respect the mechanism is closely analogous in spirit to vacuum-energy sequestering~\cite{Kaloper:2013zca,Kaloper:2014dqa,Kaloper:2014fca,Kaloper:2015jra,Kaloper:2016yfa,Kaloper:2016jsd,DAmico:2017ngr,Padilla:2018hvp,Coltman:2019mql,El-Menoufi:2019qva}: an apparently cosmological-constant-like integration constant is not fixed by a local field equation, but by an additional leafwise global constraint.

\subsection{Vacuum energy cancellation}

The principal virtue of the leafwise global constraint becomes manifest upon separating the radiatively unstable vacuum contribution from the dynamical part of the matter effective action. 
Analogous to Eq.~\eqref{Gam split z=1}, we decompose the leaf effective action as
\begin{equation}
\Gamma_{\rm m}[q_\sigma,\Psi_\sigma] = -V_{\rm vac} {\cal V}_4(\sigma)  +\Gamma_{\rm dyn}[q_\sigma,\Psi_\sigma]\,,
\label{Gamma vacuum split}
\end{equation}
where, as before,~$\Gamma_{\rm dyn}$ denotes the remainder after extracting the field-independent volume term, and is normalized to contain no constant vacuum contribution.
The intrinsic matter stress tensor correspondingly takes the form
\begin{equation}
\tensor{\tau}{^{(\rm m)a}_b} = -V_{\rm vac} \tensor{q}{^a_b} + \tensor{\tau}{^{(\rm dyn)a}_b}\,.
\label{tau vacuum split}
\end{equation}
Since~$\tau^{(\rm m)}=-4V_{\rm vac}+\tau^{(\rm dyn)}$, the three vacuum-energy contributions to Eq.~\eqref{Ein khoron effective} cancel identically: 
\begin{equation}
-\kappa^2V_{\rm vac} +2\kappa^2V_{\rm vac} -\kappa^2V_{\rm vac} =0\,.
\end{equation}
The effective Einstein equations~\eqref{Ein khoron effective} therefore become
\begin{equation}
\begin{split}
\tensor{G}{^{(4)a}_b}
= \kappa^2 \tensor{T}{^{(H)a}_b} &+ \frac{3}{2} \tensor{q}{^a_b}  \left\langle H_{cd}H^{cd} + {\cal L}_nH \right\rangle_\sigma   \\
&+\kappa^2\left( \tensor{\tau}{^{(\rm dyn)a}_b} -\frac{1}{2} \delta^a_b \big\langle \tensor{\tau}{^{(\rm dyn)}} \big\rangle_\sigma\right) + \kappa^2 \frac{\Gamma_{\rm dyn}[q_\sigma,\Psi_\sigma]}{{\cal V}_4(\sigma)} \, \delta^a_b \,.
\end{split}
\label{Ein khoron vacuum cancelled}
\end{equation}
Thus the radiatively unstable vacuum energy has disappeared completely from the gravitational field equations. Only the dynamical part of the matter effective action remains.

The cancellation may equivalently be expressed directly in terms of shifts of the renormalized matter effective action. Under the shift
\begin{equation}\label{eq:Gammashift}
\Gamma_{\rm m}\longrightarrow \Gamma_{\rm m} -\Delta V{\cal V}_4(\sigma)\,,
\end{equation}
the intrinsic stress tensor and the average of~$\Xi$ transform as
\begin{equation}
\tensor{\tau}{^{(\rm m)a}_b} \longrightarrow \tensor{\tau}{^{(\rm m)a}_b} -\Delta V\,\delta^a_b\,,
\qquad
\langle\Xi\rangle_\sigma \longrightarrow \langle\Xi\rangle_\sigma+\Delta V\,,
\end{equation}
and these shifts cancel in the effective Einstein equation. In fact, the transformation extends to the local multiplier profile as the khoron equation~\eqref{khoron eom for Xi} is invariant under
\begin{equation}
\Xi(X)\longrightarrow\Xi(X)+\Delta V\,.
\end{equation}
The shift of the leaf average displayed above is therefore the global part of a corresponding shift of the complete multiplier field. In this sense,~$\Xi$ is the covariant carrier of the global response that was implemented by the projectable lapse in the foliation-based formulation. Equivalently, Eq.~\eqref{f average Xi} shows that the flux integration constant shifts as
\begin{equation}\label{eq:fshift}
    f^2\rightarrow f^2-2\kappa^2\Delta V\,.
\end{equation}
Thus a change in the renormalized vacuum energy maps one solution into another with a correspondingly shifted five-form integration constant (provided $f^2-2\kappa^2\Delta V\geq 0$; c.f.~the discussion on page \pageref{Damienfootnote}), while leaving the intrinsic gravitational equations unchanged. Notice also how the same shifts \eqref{eq:Gammashift}--\eqref{eq:fshift} constitute a symmetry of the 5d Einstein equations \eqref{5d Einstein Xi}.

We end this section by comparing this set-up to unimodular gravity~\cite{vanderBij:1981ym,Buchmuller:1988wx,Buchmuller:1988yn,Padilla:2014yea}. In unimodular gravity, restricting the metric determinant yields the traceless Einstein equations. The Bianchi identity then restores the cosmological constant as an undetermined integration constant. By itself, however, unimodular gravity supplies no additional equation that fixes this constant in a radiatively stable way. If its value is determined by requiring a small Ricci scalar, or equivalently a small effective cosmological constant, on some late-time hypersurface, that condition simply selects the required value of the radiatively sensitive integration constant after the matter vacuum energy is known.

The situation here is different. The four-form field equation likewise introduces an integration constant~$f$, but the theory contains an additional leafwise global equation: the einbein equation~\eqref{rho equation avg}, which relates the leafwise average of the normal multiplier directly to the complete renormalized matter effective action. Together with the normal Einstein constraint and the requirement that the five-form solution contain a single constant~$f$, this fixes the value of~$f^2$ that enters the tangential gravitational equations. A shift of vacuum energy therefore changes the global constraint at the same time as it changes the matter stress tensor, and the two effects cancel without imposing a new curvature boundary condition. Put differently, the crucial ingredient beyond unimodular gravity is not merely the appearance of an integration constant, but the presence of an independent global equation that determines its gravitational contribution in a manner insensitive to constant shifts of the matter effective action.

\section{Bulk matter dynamics and Casimir energies}
\label{Casimir sec}

Thus far, matter has been assumed to be ultra-local along the compact direction. In the foliation language of the previous sections, each four-dimensional leaf therefore carries an independent copy of the matter theory; equivalently, in the covariant khoron formulation of Sec.~\ref{sec:stuck}, the matter functional contains no derivatives connecting neighbouring leaves. This assumption ensures that matter loops generate only contributions local in the extra dimension. On the backgrounds of interest, vacuum-energy corrections are therefore extensive in its proper length,
\begin{equation}
L=\int^\ell_0 \dd y\,N(y)\,, \qquad y\sim y+\ell \,,
\end{equation}
and hence have precisely the lapse dependence removed by the global constraint.

It is natural to ask what happens beyond this ultra-local approximation. Once matter fields acquire gradient terms along the compact direction, neighbouring leaves become coupled and matter can propagate around the circle. Quantum loops can then probe its global topology and generate finite Casimir energies. Unlike ordinary local vacuum-energy corrections, these contributions have a non-trivial dependence on~$L$ and are therefore not automatically cancelled by our mechanism. We now quantify this effect and identify a simple spectral condition under which it is suppressed.

For simplicity, consider a product background~$\mathcal M_4\times S^1$,
\begin{equation}
\dd s^2=N^2(y)\dd y^2+g_{\mu\nu}(x)\dd x^\mu \dd x^\nu\,,
\end{equation} 
and a collection of matter fields whose quadratic operators take the Laplace form 
\begin{equation}
\Delta_i = -(\nabla_4^2+E_i) -\frac{s_i^2}{N}\partial_y \left(\frac{1}{N}\partial_y \right) +m_i^2\,.
\label{eq:matter-Laplace}
\end{equation}
Here~$i$ runs over bosons, fermions, and ghosts, while~$s_i$ parametrizes the strength of propagation along the compact direction. Moreover,~$E_i$ is the endomorphism appearing in the four-dimensional Laplace-type operator. For example,~$E_s=0$ for a minimally coupled real scalar, whereas~$E_f=-\frac{1}{4} R^{(4)}$ for a Dirac fermion, such that~$\slashed D^2=-\nabla_4^2+\frac{1}{4} R^{(4)}$. This notation allows bosons and fermions to be treated uniformly. The corresponding effective circumference seen by species~$i$ is
\begin{equation}
L_i=\frac{L}{s_i}\,.
\end{equation}
For the species included in the following Kaluza–Klein analysis we take~$s_i>0$; strictly ultra-local fields are instead described by the preceding sections.
Allowing for a possible twist~$\alpha_i$ in the boundary condition, the Kaluza–Klein eigenvalues are
\begin{equation}
\lambda_{n,i} =  \left( \frac{2\pi(n+\alpha_i)}{L_i} \right)^2 \,, \qquad n\in\mathbb Z \,,
\end{equation}
with~$\alpha_i=0$ for periodic fields and, for example,~$\alpha_i=1/2$ for antiperiodic fermions. The one-loop Euclidean effective action is
\begin{equation}
\Gamma_1 = \frac{1}{2} {\rm Str}\log\Delta = -\frac{1}{2}\int_0^\infty\frac{\dd t}{t}\, {\rm Str}\,e^{-t\Delta}\,,
\label{eq:one-loop-supertrace}
\end{equation}
where the supertrace
\begin{equation}
{\rm Str}\,X = \sum_i(-1)^{F_i}n_iX_i \,,
\label{str}
\end{equation}
and~$(-1)^{F_i}$ denotes the sign with which species~$i$ enters the one-loop determinant, including the appropriate ghost contributions. Here,~$n_i$ denotes any additional species or internal multiplicity, while the trace over Lorentz/spinor indices is included in the heat-kernel coefficients below. Ghost fields are included as separate species with their appropriate statistics and multiplicities. The four-dimensional heat kernel has the usual expansion
\begin{equation}
{\rm Tr}_4\,e^{t(\nabla_4^2+E_i)} = \frac{1}{(4\pi t)^2} \int \dd^4x\sqrt g\, \sum_{p=0}^\infty t^p a_p^{(i)}\,,
\end{equation}
where
\begin{equation}
a_0^{(i)}={\rm tr}_i\,\mathbf 1_i\,, \qquad
a_1^{(i)}={\rm tr}_i\left(E_i+\frac16R^{(4)}\mathbf 1_i\right)\,,
\end{equation}
and the higher coefficients contain higher powers of the curvature~\cite{Vassilevich:2003xt}. The important step is to Poisson resum the Kaluza–Klein tower,
\begin{equation}
\sum_{n\in\mathbb Z} e^{-4\pi^2t(n+\alpha_i)^2/L_i^2} = \frac{L_i}{2\sqrt{\pi t}} \sum_{w\in\mathbb Z} e^{-L_i^2w^2/4t} e^{2\pi iw\alpha_i}\,.
\label{eq:Poisson-Casimir}
\end{equation}
This representation cleanly separates the one-loop action into
\begin{equation}
\Gamma_1=\Gamma_{\rm loc}+\Gamma_{\rm Cas}\,,
\label{eq:local-Casimir-split}
\end{equation}
corresponding respectively to the zero-winding and non-zero-winding sectors. As we will see, the first piece is harmless for the vacuum-energy mechanism, whereas the second captures the qualitatively new, topology-sensitive effect of bulk propagation.

The~$w=0$ sector is insensitive to the compact topology and coincides with the local contribution obtained when the compact direction is decompactified.
In dimensional regularization, its first terms are
\begin{align}
\Gamma_{\rm loc} = \int \dd^4x\sqrt g\,L \left[ \frac{1}{120\pi^2} {\rm Str}\left(\frac{a_0^{(i)}m_i^5}{s_i}\right) - \frac{1}{48\pi^2} {\rm Str}\left( \frac{a_1^{(i)}m_i^3}{s_i} \right) +\cdots \right]\,.
\label{eq:local-one-loop}
\end{align}
When adopting a cutoff scheme, the same sector also contains the usual power-law ultraviolet divergences which renormalize local five-dimensional operators. Most importantly, every term in~$\Gamma_{\rm loc}$ is proportional to the proper circumference~$L$. The leading term is therefore simply a renormalization of the local vacuum energy already considered in the previous sections. The curvature-dependent terms renormalize the gravitational couplings and generate higher-derivative operators. None of these local radiative corrections changes the basic vacuum-energy cancellation: they merely renormalize coefficients of local operators already allowed by the EFT and retain the requisite extensive dependence on the compact direction.

The genuinely new contribution comes from~$w\neq 0$. These winding sectors describe virtual particles that propagate around the compact direction and are therefore sensitive to its global size. The factor $e^{-L_i^2w^2/4t}$ suppresses the short-proper-time region, making the resulting Casimir contribution ultraviolet finite and regulator independent. Explicitly,
\begin{equation}
\Gamma_{\rm Cas} = -\frac{1}{\sqrt\pi(4\pi)^2} \int \dd^4x\sqrt g \sum_{p=0}^{\infty} \sum_{w=1}^{\infty} {\rm Str} \Bigg[ \frac{L}{s_i}\, a_p^{(i)} \cos(2\pi w\alpha_i) \left( \frac{Lw}{2s_im_i} \right)^{p-5/2} K_{p-5/2} \left( \frac{m_iLw}{s_i} \right) \Bigg]\,.
\label{Gammacas}
\end{equation}
For fields light compared with the compactification scale,~$m_iL_i\ll1$, this gives, for periodic boundary conditions,
\begin{align}
\Gamma_{\rm Cas} = -\int \dd^4x\sqrt g \left[ \frac{3\zeta(5)}{4\pi^2L^4} {\rm Str}\left(s_i^4a_0^{(i)}\right) + \frac{\zeta(3)}{8\pi^2L^2} {\rm Str}\left(s_i^2a_1^{(i)}\right) +\cdots \right]\,.
\label{eq:leading-casimir-terms}
\end{align}
The ellipsis denotes curvature and mass corrections; sufficiently high orders in a na\"ive small-mass expansion require some care because of infrared sensitivity. The full expression~\eqref{Gammacas} remains the appropriate starting point when such infrared-sensitive terms become important.

Equation~\eqref{eq:leading-casimir-terms} makes clear why bulk propagation of matter is qualitatively different from the ultra-local theory. Schematically,
\begin{equation}
\Gamma_{\rm Cas} = -\int \dd^4x\sqrt g \left[ \frac{C_0}{L^4} + \frac{C_1R^{(4)}}{L^2} +\ldots \right]\,.
\label{eq:Casimir-schematic}
\end{equation}
The curvature-dependent term represents a fractional correction of order~$(M_{\rm Pl}L)^{-2}$ to the Einstein-Hilbert term, and higher-curvature terms are similarly suppressed when~$R^{(4)}L^2\ll1$. The leading~$L^{-4}$ term is qualitatively different: it is a genuine four-dimensional vacuum energy and is not Planck-suppressed.

Returning to Lorentzian signature, there is a simple way to see directly why its non-extensive dependence on~$L$ matters. Consider a generic vacuum contribution
\begin{equation}
\Gamma_{\rm m}^{\rm vac} = - \int \dd^4x\sqrt{-g}\,U(L) \,.
\label{eq:generic-UL}
\end{equation}
Its contribution to the effective four-dimensional Einstein equation derived in Sec.~\ref{sec:def} is proportional to
\begin{equation}
\left[ \frac{U(L)}{L}-U'(L) \right]\,\delta^\mu_{\nu}\,.
\label{eq:UL-cancellation}
\end{equation}
It therefore cancels precisely when~$LU'(L)=U(L)$, {\it i.e.}, when~$U(L)\propto L$. This includes all local bulk vacuum-energy contributions. By contrast,~$U(L)\propto L^{-4}$ for the leading Casimir term, which leaves a non-vanishing gravitational source.

Consequently, if matter is allowed to propagate along the compact direction, the leading Casimir energy must either be phenomenologically negligible or be suppressed by the spectrum. For light periodic fields, cancellation of the leading~$L^{-4}$ contribution requires
\begin{equation}
{\rm Str}\left(s_i^4a_0^{(i)}\right)=0\,.
\label{eq:Casimir-supertrace}
\end{equation}
After the usual cancellation of unphysical gauge and ghost degrees of freedom, this amounts to a constraint on the weighted bosonic and fermionic spectrum, but it does not require supersymmetry.
This condition by itself need not cancel subleading mass-dependent terms; cancellation at finite masses or for general boundary conditions instead requires the appropriate cancellation of the full winding sum~\eqref{Gammacas}.

This situation is reminiscent of the spectral cancellations encountered in misaligned-supersymmetry constructions~\cite{Abel:2015oxa}. There, cancellations among bosonic and fermionic states can suppress the cosmological constant without level-by-level supersymmetric degeneracy. The analogy here is limited but useful: our mechanism removes the extensive, local vacuum-energy contribution through the gravitational and flux constraints, while the finite, topology-sensitive Casimir piece must instead be controlled by the spectrum (or be sufficiently small phenomenologically).

This perspective also clarifies the status of the ultra-local approximation used throughout the preceding sections. It is not required for the cancellation of local radiative corrections; rather, it prevents matter loops from probing the compact direction and thereby generating non-extensive dependence on its circumference. Once matter propagates in the bulk, finite winding contributions constitute the leading new obstruction.

Importantly, keeping matter ultra-local does not eliminate this issue altogether once graviton loops are included. Unlike the matter sector, gravity necessarily propagates along the compact direction in the~$z=1$ theory of Sec.~\ref{sec:def}. Quantum gravitational fluctuations therefore also possess non-zero winding sectors. Their zero-winding contribution merely renormalizes local bulk operators and is therefore compatible with the cancellation mechanism, whereas their finite winding contribution generates a genuine gravitational Casimir energy of order~$L^{-4}$. Gauge-fixing ghosts remove the unphysical gravitational polarizations but do not in general cancel the contribution of the physical graviton modes. Consequently, once graviton loops are included, the relevant spectral cancellation must involve the complete set of propagating bulk degrees of freedom, including gravity, rather than the matter sector alone. This represents an additional condition on the bulk spectrum, beyond the vacuum-energy cancellation mechanism itself. The essential distinction remains the same: local bulk vacuum energy is removed by the constraint mechanism, while finite topology-sensitive winding contributions require additional control.

\section{Discussion} \label{sec:discuss}

In this paper we have extended the framework proposed in~\cite{Khoury:2026eqr} beyond the ultra-local~$z=0$ limit by including the leading~$z=1$ dynamics along a compact extra dimension. Within the semiclassical truncation considered in the main construction, gravity and the higher-form sector are treated classically and matter is taken to be ultra-local in the compact direction. At the level of the background equations, the cancellation of the radiatively unstable matter vacuum energy persists. The projectable lapse continues to provide a constraint that is global over each four-dimensional slice, while the higher-form flux adjusts so that constant shifts of the renormalized matter effective action drop out of the effective gravitational equations. For maximally symmetric vacuum configurations, periodicity of the compact dimension then enforces vanishing four-dimensional curvature. 

The inclusion of bulk gravitational dynamics nevertheless imposes a non-trivial consistency requirement. Generic ${z=1}$ extrinsic-curvature couplings produce a non-Fierz–Pauli mass term for the non-zero Kaluza–Klein gravitons and an additional scalar ghost. At quadratic order the healthy branch satisfies~$\lambda=\mu>0$, while the five-dimensional Einstein–Hilbert theory corresponds to the special point~$\lambda=\mu=1$. This observation motivated the covariant formulation developed in Sec.~\ref{sec:stuck}, in which the gravitational and top-form sectors are fully five-dimensional and diffeomorphism covariant, while a spacelike `khoron' scalar dynamically defines the preferred foliation. Projectability is enforced by an auxiliary einbein on the one-dimensional space of khoron leaves together with the multiplier~$\Xi$. In this language the global character of the original lapse constraint is not eliminated but isolated in the leaf-space einbein equation. Together with the normal Einstein constraint and the fact that the five-form flux is characterized by a single spacetime constant, this global equation determines the flux contribution to the intrinsic Einstein equations. A constant shift of the renormalized matter vacuum energy shifts~$\Xi$ and the five-form integration constant accordingly, while leaving the intrinsic gravitational equations unchanged.

Allowing matter itself to propagate around the compact direction reveals an important limitation, and at the same time clarifies precisely what the mechanism accomplishes. The zero-winding part of the quantum effective action is local from the five-dimensional perspective. Its vacuum-energy contribution is extensive in the proper circumference~$L$, and therefore has the dependence on the projectable lapse required for cancellation; curvature-dependent terms instead renormalize local gravitational operators. Non-zero winding sectors are qualitatively different. They probe the global topology of the compact dimension and generate finite Casimir contributions with non-extensive dependence on~$L$, scaling as a four-dimensional vacuum energy of order~$L^{-4}$ for light bulk fields. Such terms do not satisfy the homogeneity condition~$LU'(L)=U(L)$ obeyed by a local bulk vacuum energy and are therefore not automatically removed by the global constraint.

Suppressing these finite contributions requires additional information about the spectrum of propagating bulk fields. In particular, for light periodic matter the leading Casimir term vanishes when~${\rm Str}(s_i^4a_0^{(i)})=0$. Once gravitational fluctuations are quantized, they too possess winding sectors and should participate in any such cancellation. This is logically distinct from the vacuum-energy mechanism itself: projectability and the flux constraint remove the local, extensive vacuum-energy operator, whereas topology-sensitive finite contributions require separate spectral control. A complete treatment of loops involving the gravitational, khoron, and constraint sectors --- and of the renormalization of the full covariant effective action --- lies beyond the matter-loop truncation studied here.

These results suggest two ingredients that a more fundamental construction would need to explain simultaneously. The first is geometric: the proper-distance measure along the compact direction is projectable, varying between leaves but not within a given four-dimensional leaf. The second is spectral: the spectrum of propagating bulk degrees of freedom must sufficiently suppress finite winding energies. More complicated compactification geometries may furnish analogues of the first ingredient, although the one-dimensional leaf-space construction used here will in general be replaced by a richer set of global moduli and constraints. Likewise, the spectral cancellations required by the Casimir sector are reminiscent of those occurring in non-supersymmetric string constructions with misaligned supersymmetry~\cite{Abel:2015oxa}, where modular invariance enforces cancellations not apparent in a low-energy field-theory description. Whether a more fundamental construction can simultaneously account for the projectable geometry and the required spectral cancellations is an interesting question for future work.

\acknowledgements
AP was supported by STFC consolidated grant number ST/T000732/1. The work of J.K. is supported in part by the DOE (HEP) Award No. DE-SC0013528. For the purpose of open access, the authors have applied a CC BY public copyright licence to any Author Accepted Manuscript version arising. \\
No new data were created during this study. 

\bibliography{ref}

@article{Blas:2010hb,
    author = "Blas, Diego and Pujolas, Oriol and Sibiryakov, Sergey",
    title = "{Models of non-relativistic quantum gravity: The Good, the bad and the healthy}",
    eprint = "1007.3503",
    archivePrefix = "arXiv",
    primaryClass = "hep-th",
    doi = "10.1007/JHEP04(2011)018",
    journal = "JHEP",
    volume = "04",
    pages = "018",
    year = "2011"
}

@article{Bellorin:2023nuh,
    author = "Bellorin, Jorge and Borquez, Claudio and Droguett, Byron",
    title = "{Quantization of the anisotropic conformal Ho{\v{r}}ava theory}",
    eprint = "2304.08646",
    archivePrefix = "arXiv",
    primaryClass = "hep-th",
    doi = "10.1103/PhysRevD.108.044035",
    journal = "Phys. Rev. D",
    volume = "108",
    number = "4",
    pages = "044035",
    year = "2023"
}

@article{Gibbons:1976ue,
  author = "Gibbons, G. W. and Hawking, S. W.",
  title = "{Action integrals and partition functions in quantum gravity}",
  journal = "Phys. Rev. D",
  volume = "15",
  year = "1977",
  pages = "2752--2756",
  doi = "10.1103/PhysRevD.15.2752"
}

@article{Duncan:1989ug,
  author = "Duncan, M. J. and Jensen, L. G.",
  title = "{Four Forms and the Cosmological Constant}",
  journal = "Nucl. Phys. B",
  volume = "336",
  year = "1990",
  pages = "100--114",
  doi = "10.1016/0550-3213(90)90344-D"
}

@article{Horava:2009uw,
  author = "Hořava, Petr",
  title = "{Quantum Gravity at a Lifshitz Point}",
  journal = "Phys. Rev. D",
  volume = "79",
  year = "2009",
  pages = "084008",
  doi = "10.1103/PhysRevD.79.084008",
  eprint = "0901.3775",
  archivePrefix = "arXiv",
  primaryClass = "hep-th"
}

@article{Horava:2009if,
    author = "Hořava, Petr",
    title = "{Spectral Dimension of the Universe in Quantum Gravity at a Lifshitz Point}",
    eprint = "0902.3657",
    archivePrefix = "arXiv",
    primaryClass = "hep-th",
    doi = "10.1103/PhysRevLett.102.161301",
    journal = "Phys. Rev. Lett.",
    volume = "102",
    pages = "161301",
    year = "2009"
}

@article{Kaloper:2013zca,
  author = {Kaloper, Nemanja and Padilla, Antonio},
  title = {Sequestering the Standard Model Vacuum Energy},
  journal = {Phys. Rev. Lett.},
  volume = {112},
  year = {2014},
  pages = {091304},
  doi = {10.1103/PhysRevLett.112.091304},
  eprint = {1309.6562},
  archivePrefix = {arXiv}
}

@article{Kaloper:2014dqa,
  author = {Kaloper, Nemanja and Padilla, Antonio},
  title = {Vacuum Energy Sequestering: The Framework and Its Cosmological Consequences},
  journal = {Phys. Rev. D},
  volume = {90},
  year = {2014},
  pages = {084023},
  doi = {10.1103/PhysRevD.90.084023},
  eprint = {1406.0711},
  archivePrefix = {arXiv}
}

@article{Kaloper:2014fca,
    author = "Kaloper, Nemanja and Padilla, Antonio",
    title = "{Sequestration of Vacuum Energy and the End of the Universe}",
    eprint = "1409.7073",
    archivePrefix = "arXiv",
    primaryClass = "hep-th",
    doi = "10.1103/PhysRevLett.114.101302",
    journal = "Phys. Rev. Lett.",
    volume = "114",
    number = "10",
    pages = "101302",
    year = "2015"
}

@article{Kaloper:2015jra,
    author = "Kaloper, Nemanja and Padilla, Antonio and Stefanyszyn, David and Zahariade, George",
    title = "{Manifestly Local Theory of Vacuum Energy Sequestering}",
    eprint = "1505.01492",
    archivePrefix = "arXiv",
    primaryClass = "hep-th",
    doi = "10.1103/PhysRevLett.116.051302",
    journal = "Phys. Rev. Lett.",
    volume = "116",
    number = "5",
    pages = "051302",
    year = "2016"
}

@article{DAmico:2017ngr,
    author = "D'Amico, Guido and Kaloper, Nemanja and Padilla, Antonio and Stefanyszyn, David and Westphal, Alexander and Zahariade, George",
    title = "{An \'etude on global vacuum energy sequester}",
    eprint = "1705.08950",
    archivePrefix = "arXiv",
    primaryClass = "hep-th",
    reportNumber = "CERN-TH-2017-115, DESY-17-080",
    doi = "10.1007/JHEP09(2017)074",
    journal = "JHEP",
    volume = "09",
    pages = "074",
    year = "2017"
}

@article{Kaloper:2016jsd,
    author = "Kaloper, Nemanja and Padilla, Antonio",
    title = "{Vacuum Energy Sequestering and Graviton Loops}",
    eprint = "1606.04958",
    archivePrefix = "arXiv",
    primaryClass = "hep-th",
    doi = "10.1103/PhysRevLett.118.061303",
    journal = "Phys. Rev. Lett.",
    volume = "118",
    number = "6",
    pages = "061303",
    year = "2017"
}

@article{Kaloper:2016yfa,
    author = "Kaloper, Nemanja and Padilla, Antonio and Stefanyszyn, David",
    title = "{Sequestering effects on and of vacuum decay}",
    eprint = "1604.04000",
    archivePrefix = "arXiv",
    primaryClass = "hep-th",
    doi = "10.1103/PhysRevD.94.025022",
    journal = "Phys. Rev. D",
    volume = "94",
    number = "2",
    pages = "025022",
    year = "2016"
}

@article{Padilla:2018hvp,
    author = "Padilla, Antonio",
    title = "{Monodromy inflation and an emergent mechanism for stabilising the cosmological constant}",
    eprint = "1806.04740",
    archivePrefix = "arXiv",
    primaryClass = "hep-th",
    doi = "10.1007/JHEP01(2019)175",
    journal = "JHEP",
    volume = "01",
    pages = "175",
    year = "2019"
}

@article{Coltman:2019mql,
    author = "Coltman, Ben and Li, Yixuan and Padilla, Antonio",
    title = "{Cosmological consequences of Omnia Sequestra}",
    eprint = "1903.02829",
    archivePrefix = "arXiv",
    primaryClass = "hep-th",
    doi = "10.1088/1475-7516/2019/06/017",
    journal = "JCAP",
    volume = "06",
    pages = "017",
    year = "2019"
}

@article{El-Menoufi:2019qva,
    author = "El-Menoufi, Basem Kamal and Nagy, Silvia and Niedermann, Florian and Padilla, Antonio",
    title = "{Quantum corrections to vacuum energy sequestering (with monodromy)}",
    eprint = "1903.07612",
    archivePrefix = "arXiv",
    primaryClass = "hep-th",
    doi = "10.1088/1361-6382/ab46f6",
    journal = "Class. Quant. Grav.",
    volume = "36",
    number = "21",
    pages = "215014",
    year = "2019"
}

@article{Sanner:2018atx,
    author = "Sanner, Christian and Huntemann, Nils and Lange, Richard and Tamm, Christian and Peik, Ekkehard and Safronova, Marianna S. and Porsev, Sergey G.",
    title = "{Optical clock comparison for Lorentz symmetry testing}",
    eprint = "1809.10742",
    archivePrefix = "arXiv",
    primaryClass = "physics.atom-ph",
    doi = "10.1038/s41586-019-0972-2",
    journal = "Nature",
    volume = "567",
    number = "7747",
    pages = "204--208",
    year = "2019"
}

@article{Lifshitz:1941a,
  author    = {Lifshitz, E. M.},
  title     = {On the Theory of Second‐Order Phase Transitions I},
  journal   = {Zh. Eksp. Teor. Fiz.},
  volume    = {11},
  pages     = {255},
  year      = {1941},
}

@article{Lifshitz:1941b,
  author    = {Lifshitz, E. M.},
  title     = {On the Theory of Second‐Order Phase Transitions II},
  journal   = {Zh. Eksp. Teor. Fiz.},
  volume    = {11},
  pages     = {269},
  year      = {1941},
}

@article{Hornreich:1975,
  author    = {Hornreich, R. M. and Luban, M. and Shtrikman, S.},
  title     = {Critical Behavior at the Onset of $ \vec{k} $‑Space Instability on the $ \lambda $ Line},
  journal   = {Phys. Rev. Lett.},
  volume    = {35},
  pages     = {1678--1681},
  year      = {1975},
}

@article{Carroll:2017gqo,
    author = "Carroll, Sean M. and Remmen, Grant N.",
    title = "{A Nonlocal Approach to the Cosmological Constant Problem}",
    eprint = "1703.09715",
    archivePrefix = "arXiv",
    primaryClass = "hep-th",
    reportNumber = "CALT-TH-2017-016",
    doi = "10.1103/PhysRevD.95.123504",
    journal = "Phys. Rev. D",
    volume = "95",
    number = "12",
    pages = "123504",
    year = "2017"
}

@article{VanNieuwenhuizen:1973fi,
    author = "Van Nieuwenhuizen, P.",
    title = "{On ghost-free tensor lagrangians and linearized gravitation}",
    doi = "10.1016/0550-3213(73)90194-6",
    journal = "Nucl. Phys. B",
    volume = "60",
    pages = "478--492",
    year = "1973"
}

@article{Hinterbichler:2011tt,
    author = "Hinterbichler, Kurt",
    title = "{Theoretical Aspects of Massive Gravity}",
    eprint = "1105.3735",
    archivePrefix = "arXiv",
    primaryClass = "hep-th",
    doi = "10.1103/RevModPhys.84.671",
    journal = "Rev. Mod. Phys.",
    volume = "84",
    pages = "671--710",
    year = "2012"
}

@article{Fierz:1939ix,
    author = "Fierz, M. and Pauli, W.",
    title = "{On relativistic wave equations for particles of arbitrary spin in an electromagnetic field}",
    doi = "10.1098/rspa.1939.0140",
    journal = "Proc. Roy. Soc. Lond. A",
    volume = "173",
    pages = "211--232",
    year = "1939"
}

@preprint{Khoury:2026eqr,
    author = "Khoury, Justin and Muntz, Benjamin and Padilla, Antonio",
    title = "{A Lapse in the Cosmological Constant Problem}",
    eprint = "2604.08659",
    archivePrefix = "arXiv",
    primaryClass = "hep-th",
    month = "4",
    year = "2026"
}

@article{ParticleDataGroup:2024cfk,
    author = "Navas, S. and others",
    collaboration = "Particle Data Group",
    title = "{Review of particle physics}",
    doi = "10.1103/PhysRevD.110.030001",
    journal = "Phys. Rev. D",
    volume = "110",
    number = "3",
    pages = "030001",
    year = "2024"
}

@article{Padilla:2014yea,
    author = "Padilla, Antonio and Saltas, Ippocratis D.",
    title = "{A note on classical and quantum unimodular gravity}",
    eprint = "1409.3573",
    archivePrefix = "arXiv",
    primaryClass = "gr-qc",
    doi = "10.1140/epjc/s10052-015-3767-0",
    journal = "Eur. Phys. J. C",
    volume = "75",
    number = "11",
    pages = "561",
    year = "2015"
}

@article{vanderBij:1981ym,
    author = "van der Bij, J. J. and van Dam, H. and Ng, Yee Jack",
    title = "{The Exchange of Massless Spin Two Particles}",
    reportNumber = "PRINT-81-0733",
    doi = "10.1016/0378-4371(82)90247-3",
    journal = "Physica A",
    volume = "116",
    pages = "307--320",
    year = "1982"
}

@article{Buchmuller:1988wx,
    author = "Buchmuller, W. and Dragon, N.",
    title = "{Einstein Gravity From Restricted Coordinate Invariance}",
    reportNumber = "DESY-88-029, ITP-UH-2/88",
    doi = "10.1016/0370-2693(88)90577-1",
    journal = "Phys. Lett. B",
    volume = "207",
    pages = "292--294",
    year = "1988"
}

@article{Buchmuller:1988yn,
    author = "Buchmuller, W. and Dragon, N.",
    title = "{Gauge Fixing and the Cosmological Constant}",
    reportNumber = "DESY-88-019, ITP-UH-1/88",
    doi = "10.1016/0370-2693(89)91608-0",
    journal = "Phys. Lett. B",
    volume = "223",
    pages = "313--317",
    year = "1989"
}

@article{Abel:2015oxa,
    author = "Abel, Steven and Dienes, Keith R. and Mavroudi, Eirini",
    title = "{Towards a nonsupersymmetric string phenomenology}",
    eprint = "1502.03087",
    archivePrefix = "arXiv",
    primaryClass = "hep-th",
    doi = "10.1103/PhysRevD.91.126014",
    journal = "Phys. Rev. D",
    volume = "91",
    number = "12",
    pages = "126014",
    year = "2015"
}

@article{Vassilevich:2003xt,
    author = "Vassilevich, D. V.",
    title = "{Heat kernel expansion: User's manual}",
    eprint = "hep-th/0306138",
    archivePrefix = "arXiv",
    doi = "10.1016/j.physrep.2003.09.002",
    journal = "Phys. Rept.",
    volume = "388",
    pages = "279--360",
    year = "2003"
}

\end{document}